\documentclass[12pt,a4paper]{iopart}
\usepackage{iopams}
\usepackage{graphicx}
\usepackage{multirow}

\newcommand{\CPCOM}{\emph{Comp. Phys. Commun.\ }}

\begin{document}

\title[Quenching dynamics: Spin-orbit coupled BECs]{Quenching dynamics of the bright solitons and other localized states in spin-orbit coupled Bose-Einstein condensates}

\author{Rajamanickam Ravisankar$^{1}$, Thangarasu Sriraman$^{1}$, Luca Salasnich$^{2,3}$ and Paulsamy Muruganandam$^{1,4}$}
\address{$^1$Department of Physics, Bharathidasan University, Tiruchirappalli 620024, Tamil Nadu, India}
\address{$^2$Dipartimento di Fisica e Astronomia “Galileo Galilei”, Universit\'a di Padova, Via Marzolo 8, 35131 Padova, Italy}
\address{$^3$Istituto Nazionale di Ottica (INO) del Consiglio Nazionale delle Ricerche (CNR), Via Nello Carrara 1, 50019 Sesto Fiorentino, Italy}
\address{$^4$Department of Medical Physics, Bharathidasan University, Tiruchirappalli 620024, Tamil Nadu, India}

\begin{abstract}
We study the dynamics of binary Bose-Einstein condensates made of ultracold and dilute alkali-metal atoms in a quasi-one-dimensional setting. Numerically solving the two coupled Gross-Pitaevskii equations which accurately describe the system dynamics, we demonstrate that the spin transport can be controlled by suitably quenching spin-orbit (SO) and Rabi coupling strengths. Moreover, we predict a variety of dynamical features induced by quenching: broken oscillations, breathers-like oscillating patterns, spin-mixing-demixing, miscible-immiscible transition, emerging dark-bright states, dark solitons, and spin-trapping dynamics. We also outline the experimental relevance of the present study in manipulating the spin states in $^{39}$K condensates.
\end{abstract}

\maketitle

\section{Introduction}

 Since the successful experimental realization of Bose-Einstein condensates (BECs) in dilute ultracold atomic gases, it becomes an excellent platform for studying a variety of physical phenomena related to quantum and condensed matter physics. For instance, the spin-orbit coupling, which couples the particle's spin and momentum, plays an important role in many condensed matter phenomena. The theoretical and the experimental works in the recent past on synthetic spin-orbit (SO) coupling in Bose and Fermi gases at ultra-low temperatures paved the way for the creation and measurement of spin Hall effect~\cite{Stanescu2008, Lin2011, Galitski2013, Zhai2015}. The physical mechanism of creating spin-orbit coupling in experiments requires electric fields of the order of trillions of volts per meter, which is inaccessible in laboratory conditions. Instead, the spin-orbit coupling can be engineered in neutral atomic BECs using laser fields. There are two types of spin-orbit coupling, namely, Rashba effect~\cite{Rashba1955}, which arises from the breaking of inversion symmetry by the induced electric field and Dresselhaus spin-orbit splitting~\cite{Dresselhaus1955} due to the lack of inversion symmetry in the material. Both the Rashba effect and Dresselhaus splitting play crucial roles in many physical phenomena including spin Hall effects~\cite{Xiao1959}, spintronics~\cite{Zutic2004}, topological insulators~\cite{Hasan2010}, quantum simulations~\cite{Johanning2009, Bloch2012}, etc. 

An interesting dynamical feature in Bose-Einstein condensates which attracted a great deal of attention is the observation of matter wave solitons. Solitons are localized wave packets that arise due to the balancing of dispersion with nonlinearity and they can propagate without changing its shape or velocity. In this context, BEC provides an excellent platform to manipulate the dispersion and nonlinearity experimentally~\cite{Morsch2006,Chin2010}. The spin-orbit coupled BECs exhibit an unusual shape-changing property of solitons which arises due to the lack of Galilean invariance~\cite{Xu2013}. The interplay of the coupling parameters and nonlinearity has been found to induce a precession of the soliton's spin under the action of an external magnetic field~\cite{Wen2016}. Further, the connection between the modulational instability and formation of soliton in spin-1/2 and spin-1 SO coupled BECs has been studied in detail~\cite{Bhat2015, Bhuvaneswari2016, GQLi2017}.

Several studies have been made by combining the linear spin-orbit and Rabi coupling between the species, and the inter- and intra-species nonlinear interactions. For instance, there are studies on the localized modes within the mean-field framework from the non-polynomial Schr{\"o}dinger equation~\cite{Luca2013, Luca2014}, stationary states and solitons in a bichromatic optical lattice~\cite{Cheng2014}, moving bright soliton dynamics~\cite{Xu2013}, matter wave bright and dark solitons and their stability properties in the semi-infinite gap of the energy spectrum~\cite{Achilleos2013-bs, Achilleos2013-ds}, flipping structural oscillation and shuttle motion of bright and stripe solitons~\cite{Sakaguchi2017}.  Further, the excitation spectrum in quasi-1D SO coupled BECs showing supersolid property has been predicted~\cite{YLi2013} and subsequently a stripe phase in the density modulation showing supersolid property, as the evidence for spontaneous long-range order in one direction, was observed experimentally~\cite{JRLi2017}.

Stationary bright solitons and a phase diagram illustrating regimes of plane wave and stripe phases in SO and Rabi coupling parameter plane both in the presence and absence of dipolar interactions are studied numerically~\cite{Chiquillo2018, Tononi2019}. Different ground state phases such as plane wave, zero-momentum, and stripes are identified besides the studies on the dynamics of dark solitons~\cite{Cao2015}. Further, the role of symmetries like parity ($\cal{P}$), time ($\cal{T}$) and spin/charge ($\cal{C}$) symmetries on the stable gap, gap-stripe solitons and nonlinear modes are reported with different traps~\cite{Kartashov2013, YZhang2015, TFXu2018b, Kartashov2014}.

Zitterbewegung oscillations have been observed experimentally by sudden quantum quenching~\cite{CQu2013}. Certain loop structures in the nonlinear dispersion relations of SO coupled BECs in the presence of weak accelerating force are identified~\cite{YZhang2019}. However, the Zitterbewegung dynamics induced by the sudden quench of SO coupling has no relevance to the nonlinear dispersion. Time evolution of the steady state condensate fraction as well as oscillating momentum distribution are investigated by time-dependent Bogoliubov-de Gennes equation with the quench of inter- and intra-species interactions and SO coupling~\cite{Deng2016}. Domain formation by the quenching process, the homogeneous and inhomogeneous Kibble-Zurek mechanism in the trapped SO coupled BECs are studied using truncated-Winger Gross-Pitaevskii (GP) equation~\cite{Liu2019}. The presence of SO coupling along with Zeeman splitting makes the coupled GP equations to be nonintegrable, which lead to certain nontrivial soliton properties, for instance, the formation of stable quasi-scalar soliton complexes~\cite{Kartashov2014-2}. The evolution of solitons in inhomogeneous gauge potential by Zeeman splitting serves as a parameter controlling the crossover between the two different integrable limits~\cite{Kartashov2019}. Moving bright solitons are also found in two-dimensional BECs with Rashba type SO coupling, which has the mobility property only in one direction up to a critical value of the velocity, beyond which delocalization occurs~\cite{HSakaguchi2014}. The existence of stable self-accelerating solitons and vortex solitons are also noticed~\cite{Qin2019}. Also, there are studies on self-trapped stable solitons that are predicted in three-dimensional free space, and vortex-bright solitons in spin-1 SO coupled BECs~\cite{CYZhang2015, SGautam2017,SGautam2018}.

We note that most of the studies focus on the properties and dynamics of solitons in quasi-1D SO coupled BECs, but only a few studies are available on quenching dynamics. In this paper, we report the dynamics of the quasi-1D pseudospin-$1/2$ system with an equal combination of Rashba coupling and Dresselhaus effect. By numerically solving the coupled Gross-Pitaevskii equations, we study the interplay between nonlinear interactions, SO, and Rabi coupling in the absence as well as in the presence of a harmonic trap.

The rest of the paper is organized as follows: In section \ref{sec:2}, we shall introduce the theoretical model with SO and Rabi couplings and the calculation of chemical potentials for the spin components. We point out the relevance of the present study for performing experiments with the help of present-day technologies in section \ref{sec:3}. In section \ref{sec:4}, we examine the symmetric behaviour as well as the dynamics of the ground state solitons in real and imaginary time evolution for different stationary states. Further, we discuss the quench dynamics of binary BEC by sudden changes of the SO and Rabi coupling strengths in section \ref{sec:5}. In section \ref{sec:6}, we report the analysis of the dynamics in the cases of plane wave and stripe phases of SO coupled BECs with Rabi coupling. Finally, we summarize our results and provide the conclusions in section \ref{sec:7}.

\section{Theoretical description and coupled Gross-Pitaevksii equations}
\label{sec:2}

In experiments, the spin-orbit coupled BECs are created, for instance, by choosing two internal spin states of $^{87}$Rb atoms within the $5S_{1/2}$, $F=1$ ground electronic manifold, which are designated as pseudo-spin up, $\vert \uparrow \rangle = \vert F = 1, m_{F}=0 \rangle$ and spin-down, $\vert \downarrow \rangle = \vert F = 1, m_{F}=-1 \rangle$~\cite{Lin2011}. Spin orbit coupling between the spin-up and spin-down states is manipulated by a pair of counter propagating Raman lasers~\cite{Stanescu2008, Lin2011, Galitski2013, Zhai2015}. 

A pseudo-spin $1/2$ Bose-Einstein condensate can be modelled by the following Hamiltonian 
\begin{equation}
\mathcal{ H} = \mathcal{ H} _{\mathrm{sp}} + \mathcal{H}_{\mathrm{int}}, 
\end{equation}
where $ \mathcal{H} _{\mathrm{sp}}$ corresponds to the single particle Hamiltonian and is given by 
\begin{equation}
\mathcal{H} _{\mathrm{sp}} = \int \Psi^* \left( \left[ \frac{ \mathbf{\hat p^2}}{2 m} + V(\mathbf{r}) \right] + \frac{\hbar \Omega}{2} \sigma_x - \frac{k_L}{m} \hat p_x \sigma_z \right) \Psi d\mathbf{r},
\end{equation}
and
\begin{equation}
 \mathcal{H}_{\mathrm{int}} = \int \left( \frac{\alpha_{\uparrow \uparrow}}{2} \vert \psi_\uparrow \vert^4 + \frac{\alpha_{\downarrow \downarrow}}{2} \vert \psi_\downarrow \vert^4 +\beta \vert \psi_\uparrow \vert^2 \vert \psi_\downarrow \vert^2 \right) d\mathbf{r} .
\end{equation}
In the above, $\Psi = \left(\psi_\uparrow, \psi_\downarrow \right)^T$, where $\psi_\uparrow$ and $\psi_\downarrow$ are the wavefunctions of the spin components, $\mathbf{\hat p} = -\mathrm{i} \hbar \left( \partial_x, \partial_y, \partial_z \right)$ is the momentum operator, $V(\mathbf r)$ is a trapping potential, $k_L$ represents the recoil wave number (strength of spin-orbit coupling) induced by the interaction with the laser beams,  $\Omega$ corresponds to the frequency of the Raman laser (Rabi mixing) which couples the two spin states, $\sigma_{x,z}$ are the $2 \times 2$ Pauli matrices, $\alpha_{jj} = 4 \pi \hbar^2 a_{jj} / m $, $(j = \uparrow, \downarrow)$ represent the intra-component interaction strengths and $\beta = 4 \pi \hbar^2 a_{\uparrow \downarrow} / m = 4 \pi \hbar^2 a_{\downarrow \uparrow} / m $, corresponds to the inter-component interaction strength and are characterized by the $s$-wave scattering lengths $a_{ij}$, $(i,j = \uparrow, \downarrow;~ i\neq j)$. Further, the wavefunctions are subjected to the normalization condition $\sum_{j = \uparrow, \downarrow} \int \vert \psi_j \vert^2 d\mathbf{r} =1$.

The pseudo-spin $1/2$ Bose-Einstein condensate with Rabi coupling can be described by a pair of coupled Gross-Pitaevskii (GP) equations as~\cite{Achilleos2013-bs}
\numparts 
\label{eq:gpsoc:1} 
\begin{eqnarray}
\mathrm{i} \partial_{t} \psi_{\uparrow} = \bigg[ -\frac{1}{2}\partial_{x}^2 & - \mathrm{i} k_L \partial_{x} + V(x) 
+ \alpha_{\uparrow \uparrow} \vert \psi_{\uparrow} \vert^2 + \beta \vert \psi_{\downarrow} \vert^2\bigg] \psi _{\uparrow} + \Omega \psi_{\downarrow}, \label{eq:gpsoc:1a} \\ 
\mathrm{i} \partial_{t} \psi_{\downarrow} = \bigg[ -\frac{1}{2}\partial_{x}^2 & + \mathrm{i} k_L \partial_{x}+ V(x) 
+ \beta \vert \psi_{\uparrow} \vert^2 + \alpha_{\downarrow \downarrow} \vert \psi_{\downarrow} \vert^2 \bigg] \psi _{\downarrow}+ \Omega \psi_{\uparrow}, \label{eq:gpsoc:1b}
\end{eqnarray}
\endnumparts 
where $V(x) = \lambda^2 x^2/2$ is the harmonic axial trap which is characterized by the parameter $\lambda$, the trap aspect ratio, $k_L \to a_{\perp} k_L$, $\Omega \to \Omega/(2 \omega_{\perp})$ are the rescaled strengths of SO and Rabi couplings, and $\psi_{\uparrow,\downarrow} = \sqrt{a_{\perp}} \psi_{\uparrow,\downarrow}$ is the rescaled spin-components wavefunction. In the above, the length is measured in units of the oscillator length, $a_{\perp} = \sqrt{\hbar/m \omega_{\perp}}$ and time is measured in units of $\omega_{\perp}^{-1}$, where $\omega_{\perp}$ is the trap frequency in the transverse direction. Further, we assume that the intra-species interaction strengths of both the components are equal, that is, $\alpha_{\uparrow \uparrow} = \alpha_{\downarrow \downarrow} = \alpha$.
One may note that, in the absence of SO and Rabi couplings, that is, with $k_L =0$ and $\Omega = 0$, the above system of coupled GP equations (\ref{eq:gpsoc:1a}) and (\ref{eq:gpsoc:1b}) reduces to the so-called Manakov model, when $\alpha, \beta<0$ and $V(x)=0$, and admits bright-bright solitons~\cite{Radhakrishnan1997, Kanna2003}. 

The chemical potentials may be deduced by assuming $\psi_{\uparrow} = \mathrm{e}^{\mathrm{i} \mu_1 t} \left( \psi_{\uparrow R} + \mathrm{i} \psi_{ \uparrow I} \right)$ and $\psi_{\downarrow } = \mathrm{e}^{\mathrm{i} \mu_2 t} \left( \psi_{\downarrow R} + \mathrm{i} \psi_{\downarrow I}\right)$. Substituting $\psi_{\uparrow, \downarrow}$ in equations (\ref{eq:gpsoc:1a}) and (\ref{eq:gpsoc:1b}) and separating real and imaginary parts, one obtains the expressions for chemical potentials $\mu_1$ and $\mu_2$ as
\numparts 
\begin{eqnarray}
\fl \mu_1 = \frac{1}{N_{\uparrow} } \int \left\{ \frac{1}{2}\left[ \left(\frac{\partial \psi_{\uparrow R}}{\partial x}\right)^2 
+ \left(V(x) + \alpha \vert \psi_\uparrow \vert^2 + \beta \vert \psi_\downarrow \vert^2 \right) \psi_{\uparrow R}^2 \right] + \left[ \Omega \psi_{\downarrow R} + k_{L} \frac{\partial \psi_{\uparrow I}}{\partial x} 
\right] \psi_{\uparrow R}\right\} dx,\nonumber \\ \label{eq:gpsoc2:mu:1} \\
\fl \mu_2 = \frac{1}{ N_{\downarrow}} \int \left\{ \left[ \frac{1}{2} \left(\frac{\partial \psi_{\downarrow R}}{\partial x}\right)^2 
+ \left( V(x) +\beta \vert \psi_\uparrow \vert^2 + \alpha \vert \psi_\downarrow \vert^2 \right) \psi _{\downarrow R}^2 \right] 
 + \left[\Omega \psi_{\uparrow R} - k_{L} \frac{\partial \psi_{\downarrow I}}{\partial x} \right] \psi_{\downarrow R} \right\} dx, \nonumber \\ \label{eq:gpsoc2:mu:2}
\end{eqnarray}
\label{eq:gpsoc2:mu}
\endnumparts 
where $N_{\uparrow} = \int \psi_{\uparrow R}^2dx$ and $N_{\downarrow} = \int \psi_{\downarrow R}^2 dx$. 

\section{Possible experimental realization and parametrization}
\label{sec:3}
The static and dynamical properties of SO coupled BECs can be described by the coupled Gross-Pitaevksii equation (4). The theoretical description in Ref.~\cite{Lin2011} is for the experimentally observed $^{87}$Rb SO coupled BECs. However, the model proposed in Ref.~\cite{Lin2011} is also applicable for other SO coupled BECs. In the present study, we considered SO coupled BECs of $^{39}$K atoms. This particular choice is mainly due to their wide tunability region, which includes both attractive as well as strong repulsive atomic interaction strengths.

The choice of parameters for the numerics in this paper are in-line with experiment constrains. For the present study, we consider the case of $^{39}$K condensate of about $2 \times 10^3$ atoms confined in the trapping potential with frequencies $\omega_{x} = 2\pi \times 81.25$\,Hz, $\omega_{\perp} = 2\pi \times 1625$\,Hz along $x$- and perpendicular directions, respectively (trap aspect ratio, $\lambda = 0.05$ and oscillator length, $a_\perp \sim 1\,\mu\mbox{m}$)~\cite{Jin2014}. The two internal hyperfine states, $\vert F= 1, m_F = -1\rangle$ and $ \vert F= 1, m_F = 0\rangle$, may be chosen and regarded as pseudo-spin up $\vert \uparrow \rangle$, and pseudo-spin down $\vert \downarrow \rangle$ states. These two spin states are populated with equal number of atoms, and their intra- and inter-species interaction strengths can be controlled by tuning $s$-wave scattering lengths through Feshbach resonance~\cite{Thalhammer2008, Papp2008}. By suitably changing the magnetic field, the $s$-wave scattering length may be varied to a wider range from $-33 a_0$ to $180 a_0$ ($a_0$ is the Bohr radius)~\cite{Jin2014, Roati2007}, and this gives the range of dimensionless interactions strengths from $-6.98$ to $38$. We also consider the contact interaction strengths, $\alpha$ and $\beta$, so as to justify the validity of quasi-condensate within the mean-field description~\cite{Luca2004, Luca2005}. See \ref{app:a} for more details. Further, if the two spin states are coupled by Raman lasers with frequency ranging from $ 2 \pi \times 32.50$\,Hz to $ 2 \pi \times 3.25$\,kHz, then the resultant dimensionless Rabi coupling strength will be of the order of $0.01$ to unity. The SO coupling term, $\sqrt{2} \pi/\lambda_{L}$, where $\lambda_{L}$ is laser wavelength, that arises from the laser geometry can be independently varied~\cite{Lin2011}. The choice of laser wavelength typically ranging from far-infra-red (FIR) to visible region could lead to the dimensionless SO coupling strength $k_L$ to lie in the domain~($0.1, 2.0$). Very recently, a generic and experimentally feasible scheme for varying the SO coupling strength to a wider range was proposed~\cite{Luo2019}.

\section{Dynamics of the Bose-Bose bright soliton}
\label{sec:4}
In the following section, we shall explore the dynamics of bright soliton in spin-orbit coupled BECs by numerically solving the GP equations (\ref{eq:gpsoc:1a}) and (\ref{eq:gpsoc:1b}) using the split-step Crank-Nicholson method~\cite{Muruganandam2009, Vudragovic2012}. The numerical simulations are carried out with a space step $dx = 0.025$ in the domain $x \in (-51.2, 51.2)$ or $x \in (-76.8, 76.8)$ with periodic boundary conditions. The stationary states are obtained using imaginary time propagation, while the dynamics are studied using real time propagation both with a time step $dt = 6.25 \times 10^{-4}$. %
To start with, we assume Gaussian wave profiles of the form
\begin{equation}
\Psi(0) = 
\left\{
\begin{array}{ll}
\displaystyle \frac{1}{\sqrt{2 \pi}} \left(1,\pm 1\right)^{T} \exp\left[-\frac{x^{2}}{2}\right] & \textrm{for} \;\; \lambda = 0, \\
\displaystyle \frac{\sqrt{\lambda}}{\sqrt{2 \pi}} \left(1,\pm 1\right)^{T} \exp\left[-\frac{x^{2}}{2}\right] & \textrm{for} \;\; \lambda \neq 0, 
\end{array}
\right.
\label{eq:Gauss-notrap}
\end{equation}%
as the initial conditions in the imaginary time propagation. %
First we create a set of initial wave profiles using imaginary time propagation by fixing the parameters as $\alpha = \beta = -0.8$, $k_L = 0.5$ and $\Omega = 0.5$. Then we study the dynamics by evolving these initial profiles in real time propagation. %
\begin{figure}[!ht]
\begin{center}
\includegraphics[width=0.99\linewidth]{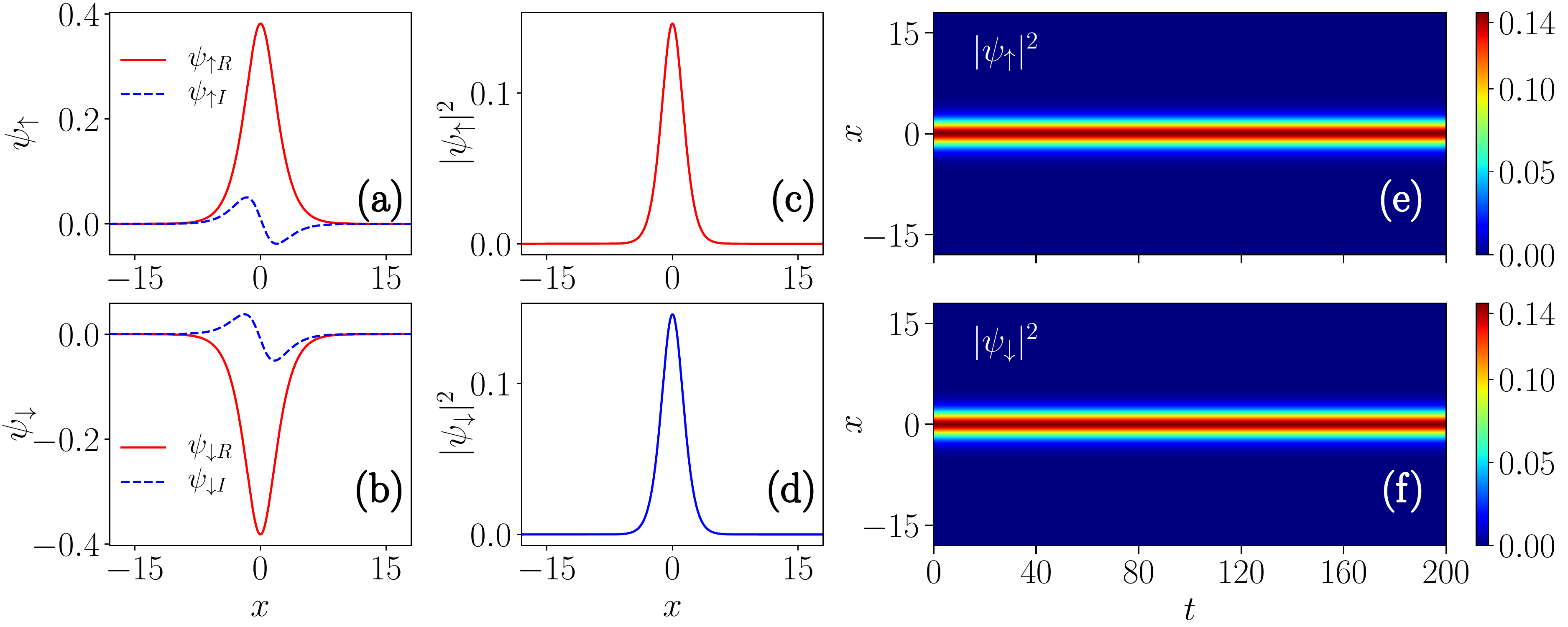}
\end{center}
\caption{Plots of the real (solid red line) and the imaginary (dashed blue line) parts of the anti-symmetric initial wavefunctions of the spin components (a) $\psi_{\uparrow}$ and (b) $\psi_{\downarrow}$, and the corresponding single soliton profiles, (c) $\vert\psi_{\uparrow}\vert^2$ and (d) $\vert \psi_{\downarrow}\vert^2$ for $\lambda = 0$, $\alpha = \beta = -0.8$, $k_L = 0.5$ and $\Omega = 0.5$. Figures (e) and (f) show the time evolution obtained from the numerical solution of the coupled GP equations~(\ref{eq:gpsoc:1a}) and (\ref{eq:gpsoc:1b}). The initial profiles are generated with the above parameters from imaginary time propagation and correspond to $\mu_{\uparrow} = \mu_{\downarrow} = -0.623$.}
\label{fig1}
\end{figure}%
For the above choice of parameters and with the normalization condition, one can identify two different states corresponding to the chemical potentials $\mu_1 = \mu_2 = -0.623$ and $-0.537$, respectively. We find certain interesting phenomena which depends on the symmetry properties of the intitial wavefunctions. For instance, if we take initial wavefunctions as an anti-symmetry form, that is, $\psi_{\uparrow}(x) = - \psi_{\downarrow}(-x)$ as shown in  figures~\ref{fig1}(a) and \ref{fig1}(b), during the time evolution, the one-soliton profiles of the spin components are preserved which is evident from  figures~\ref{fig1}(c), \ref{fig1}(d),  figures~\ref{fig1}(e) and \ref{fig1}(f) as they remain stationary during the time evolution. However, if we take the initial wavefunctions as a cross-symmetry form, that is, $\psi_{\uparrow}(x) = \psi_{\downarrow}(-x)$ as shown in  figures~\ref{fig2}(a) and \ref{fig2}(b), the corresponding two soliton profiles, shown in  figures~\ref{fig2}(c) and \ref{fig2}(d), become dynamically unstable and tend to move along $+x$ and $-x$ directions during time evolution as illustrated in  figures~\ref{fig2}(e) and \ref{fig2}(f). Actually, this type of soliton complexes arises due to the effect of SO coupling~\cite{Kartashov2014-2}. %
\begin{figure}[!ht]
\begin{center}
\includegraphics[width=0.99\linewidth]{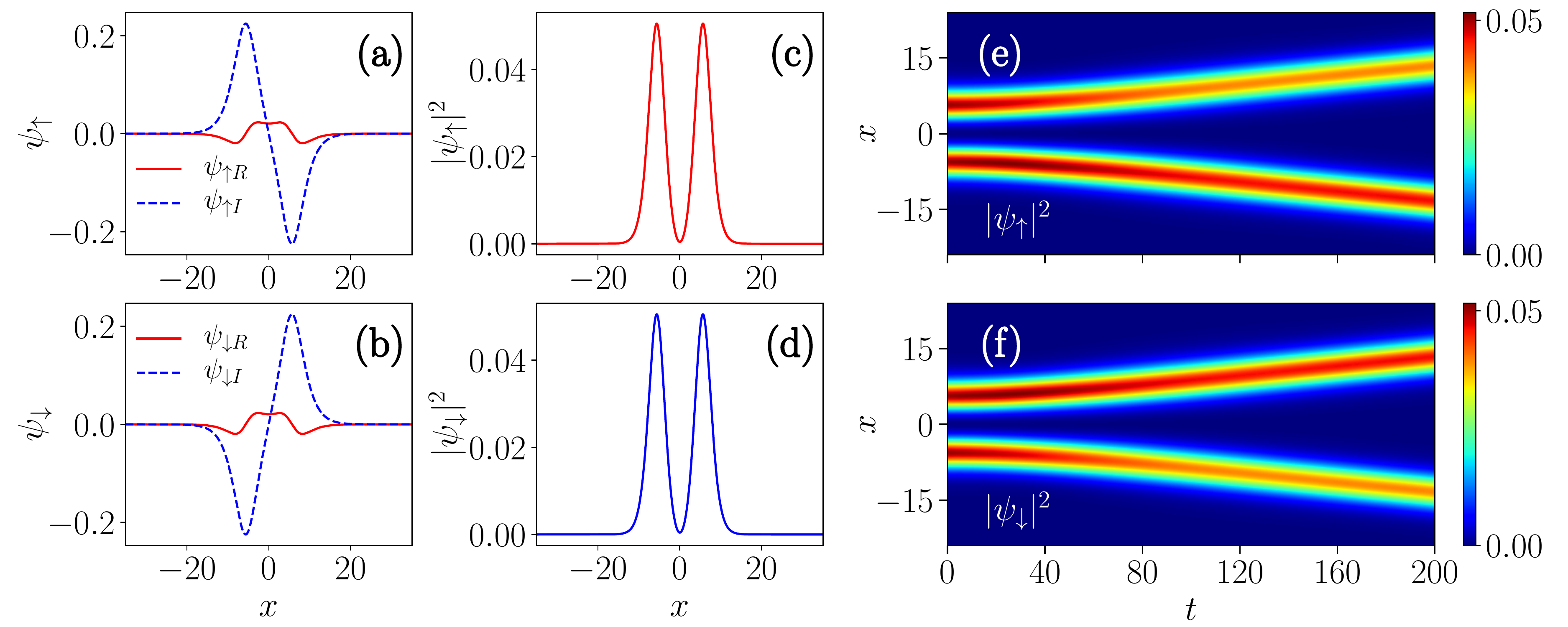}
\end{center}
\caption{Plots of the real (solid red line) and the imaginary (dashed blue line) parts of the cross-symmetric initial wavefunctions, (a) $\psi_{\uparrow}$ and (b) $\psi_{\downarrow}$, and the corresponding two-soliton density profiles (c) $\vert\psi_{\uparrow}\vert^2$ and (d) $\vert \psi_{\downarrow}\vert^2$ for the set of parameters as given in figure \ref{fig1} with $\mu_{\uparrow} = \mu_{\downarrow} \approx -0.537$. Time evolution of the densities (c) $\vert\psi_{\uparrow}\vert^2$ and (f) $\vert \psi_{\downarrow}\vert^2$ showing the unstable nature as they move in the $\pm x$ direction.}
\label{fig2}
\end{figure}%
In general, solitons with opposite phases repel each other and vice versa when they are in-phase~\cite{Ostrovskaya1999}. In the present case, the solitons are in opposite phase and therefore they do not attract each other.

We have also studied the dynamics of stripe solitons that exist for larger values of $k_L$ and $\Omega$. For the parameters $\lambda=0$, $\alpha = \beta = -0.8$, $k_L = 4$ and $\Omega = 2$, the anti-symmetric initial wavefunction is stable, whereas the cross-symmetric initial wavefunction is unstable, this is manifested in their respective chemical potentials $-8.207$ and $-8.148$ with anti-symmetric initial wavefunction having the lowest chemical potential.
\begin{figure}[!ht]
\begin{center}
\includegraphics[width=0.99\linewidth]{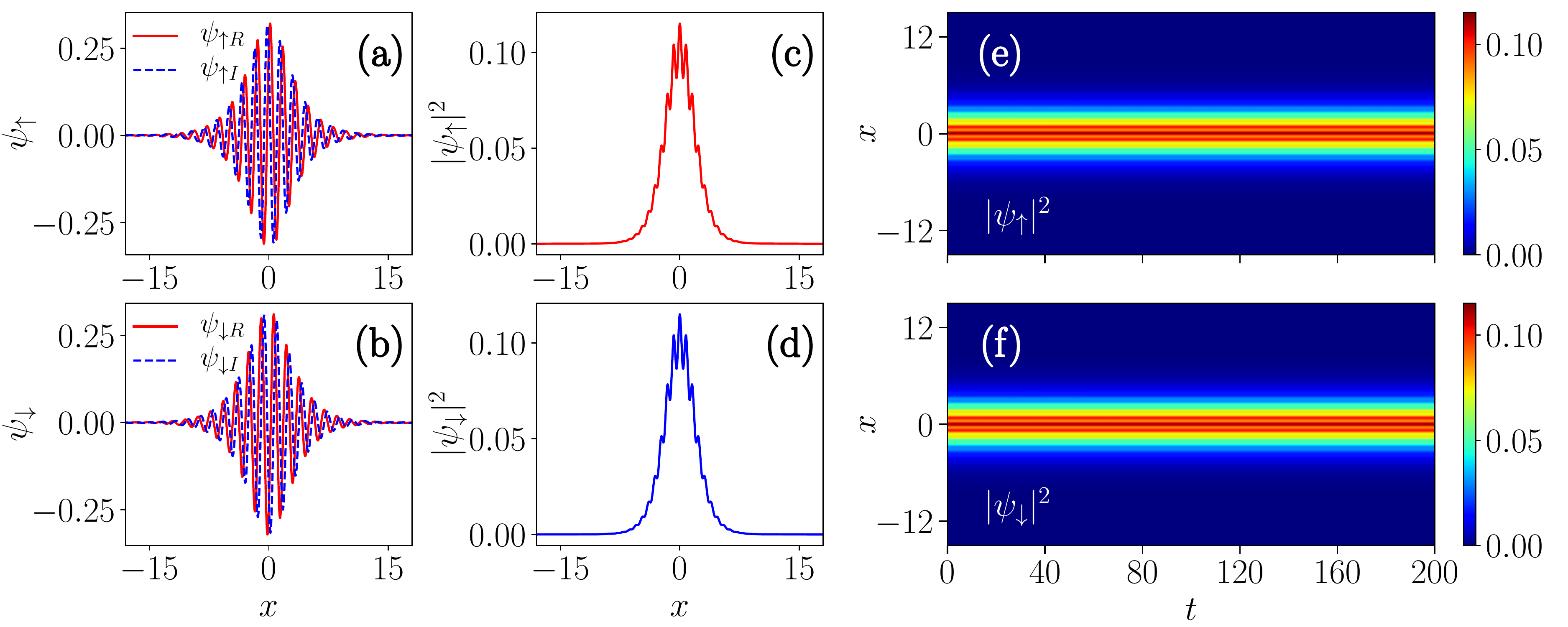}
\end{center}
\caption{Plots showing the real (solid red line) and the imaginary (dashed blue line) parts of the anti-symmetric initial wavefunctions, (a) $\psi_{\uparrow}$ and (b) $\psi_{\downarrow}$, and the density profiles of the stripe solitons, (c) $\vert\psi_{\uparrow}\vert^2$ and (d) $\vert \psi_{\downarrow}\vert^2$, for $\lambda = 0$, $\alpha = \beta = -0.8$, $k_L = 4$ and $\Omega = 2$ with $\mu_{\uparrow} = \mu_{\downarrow} = -8.207$. The time evolution of the densities, (e) $\vert \psi_{\uparrow}\vert ^2$, and (f) $\vert \psi_{\downarrow}\vert ^2$, confirming stationary nature of the stripe solitons.}
\label{fig3}
\end{figure}%
\begin{figure}[!ht]
\begin{center}
\includegraphics[width=0.99\linewidth]{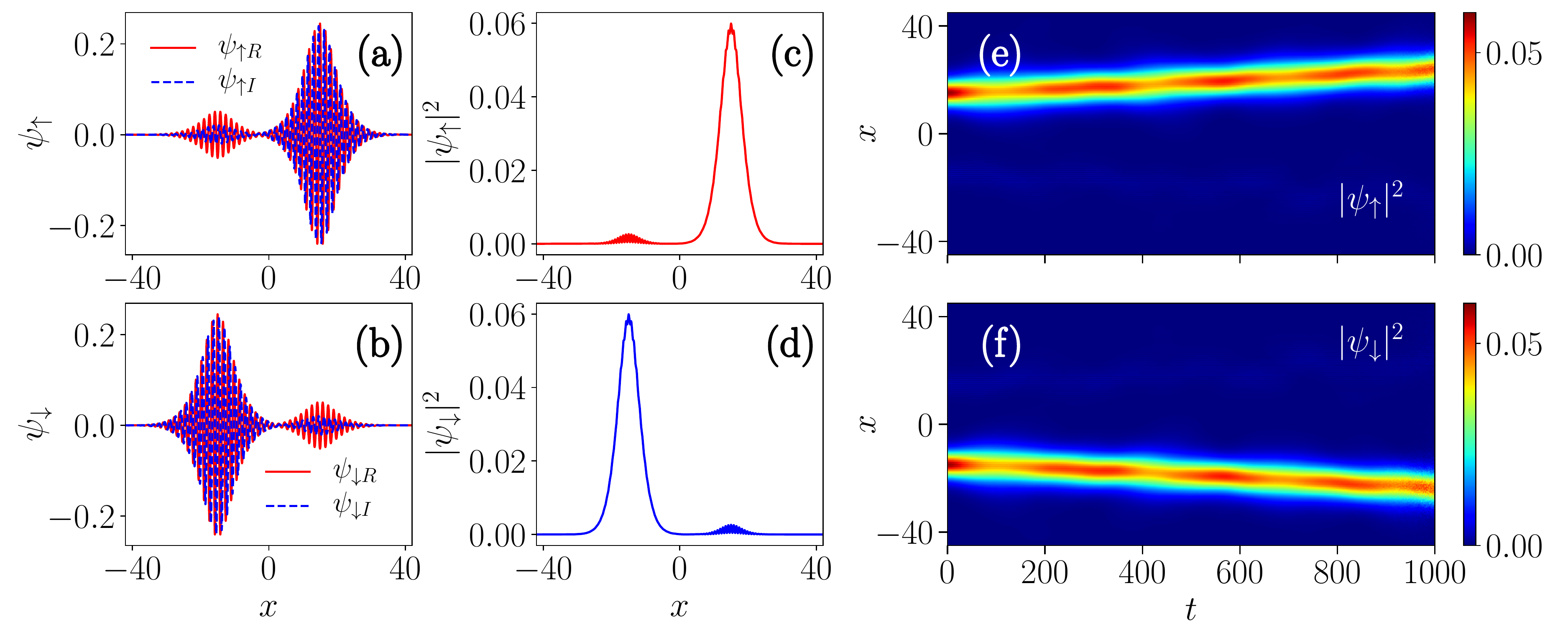}
\end{center}
\caption{Plots of the the real (solid red line) and the imaginary (dashed blue line) parts of the cross-symmetric initial wavefunctions, (a) $\psi_{\uparrow}$ and (b) $\psi_{\downarrow}$, and the initial density profiles, (c) $\vert\psi_{\uparrow}\vert^2$ and (d) $\vert \psi_{\downarrow}\vert^2$, for $\lambda = 0$, $\alpha = \beta = -0.8$, $k_L = 4$, and $\Omega =2$ with $\mu_{\uparrow} = \mu_{\downarrow} \approx -8.148$. The time evolution of stripe solitons, (e) $\vert \psi_{\uparrow}\vert ^2$, and (f) $\vert \psi_{\downarrow}\vert ^2$, showing the propagation of the spin component densities in $\pm x$ directions.}
\label{fig4}
\end{figure}%
Figures~\ref{fig3}(a) - \ref{fig3}(d) display the stable anti-symmetric wavefunctions of the spin components, and the corresponding density profiles, $\vert \psi_{\uparrow}\vert ^2$ and $\vert \psi_{\downarrow}\vert ^2$, of the stripe solitons. During time evolution, the stripe solitons emerging from the anti-symmetric initial wavefunction remain stationary [see figure \ref{fig3}(e) and \ref{fig3}(f)] whereas the stripe solitons emerging from the cross-symmetric initial wavefunctions propagate in the $\pm x$ direction as shown in  figures~\ref{fig4}(e) and \ref{fig4}(f).

The plane and stripe wave phases discussed above arise from the single-particle dispersion spectrum~\cite{Achilleos2013-bs, YLi2012} which is given by dispersion relation, 
\begin{equation}
\label{eq:sps}
\omega_{\pm}(k_x) = \frac{1}{2}k_x^2 \pm \sqrt{k_{L}^2 k_x^2 + \Omega^2}. \label{eq:disperse}
\end{equation}
This dispersion relation can be derived from equations (\ref{eq:gpsoc:1a}) and (\ref{eq:gpsoc:1b}) using the plane wave solutions $\psi_{\uparrow, \downarrow} = \phi_{\uparrow, \downarrow} \exp[\mathrm{i}(k_x x - \omega t)]$, where, $\phi_{\uparrow, \downarrow}\ll 1$, is the ground state amplitude. %
Further, it is interesting to point out that the bright solitons described in  figures~\ref{fig1} and \ref{fig2} exist only for $ \Omega > k_L^2$, while the stripe solitons shown in  figures~\ref{fig3} and \ref{fig4} arise when $\Omega < k_L^2$, as emerged from the single particle dispersion-spectrum~\cite{Achilleos2013-bs,Ravisankar2020}. A phase diagram corresponding to the dispersion relation (\ref{eq:disperse}) may be found in Ref.~\cite{Ravisankar2020}.

Besides, it will be worth mentioning about the symbiotic stripe solitons, which one can expect in the case of repulsive intra- and attractive inter-species interactions~\cite{Gracia2005, Adhikari2005}. Interestingly, we identify symbiotic solitons in the SO coupled BECs with repulsive intra-species interactions of unequal strengths, that is $\alpha_{\uparrow\uparrow} \neq \alpha_{\downarrow\downarrow}$, $\alpha_{\uparrow\uparrow} > 0$, $\alpha_{\downarrow\downarrow} > 0$ and $\beta < 0$ in equation (4). See Appendix B for more details.

\section{Dynamics of binary and spin-orbit coupled BEC{s} by quenching spin-orbit and Rabi coupling parameters}
\label{sec:5}

Next, we study the quench induced dynamics by applying sudden changes in the SO and Rabi couplings. It may be noted that equations (\ref{eq:gpsoc:1a}) and (\ref{eq:gpsoc:1b}) break the Galilean invariance, as a consequence of the spin-orbit coupling term. The dynamical properties are considerably different from that of typical binary BECs, which do not have spin-orbit coupling~\cite{Xu2013, Zhu2012}. Hence, it is of great interest to explore the dynamics. For this purpose, we first prepare the stationary solution of the coupled GP equations in the imaginary time propagation by setting $k_L = \Omega = 0$. This stationary state is then evolved in the realtime propagation, during which the spin-orbit and Rabi coupling parameters are introduced. 

In the case of attractive inter- and intra-component interactions, the introduction of the SO and Rabi coupling leads to periodic oscillations and decay and revival of the spin components. On the other hand, while with the repulsive interactions, the spin densities exhibit temporal oscillations, spin-mixing and demixing, filament formation, dark solitons, and spin-flipping dynamics.

\subsection{Quenching dynamics of the Bose-Bose bright solitons in spin-orbit coupled BECs with attractive interactions}
\label{sec:5:1}

First, we consider the case of attractive BEC and study the dynamics by switching on the spin-orbit and/or Rabi coupling strengths. We prepare the stationary profile of the wavefunctions with the help of imaginary-time propagation by fixing the interaction strengths $\alpha = \beta = -0.8$ and by setting the SO and Rabi coupling strengths to zero. Then this stationary profile is evolved and during the time evolution, the spin-orbit (or both SO and Rabi) coupling is turned on at a finite time, say for instance, $t = 10$. 

\begin{figure}[!ht]
\begin{center}
\includegraphics[width=0.99\linewidth]{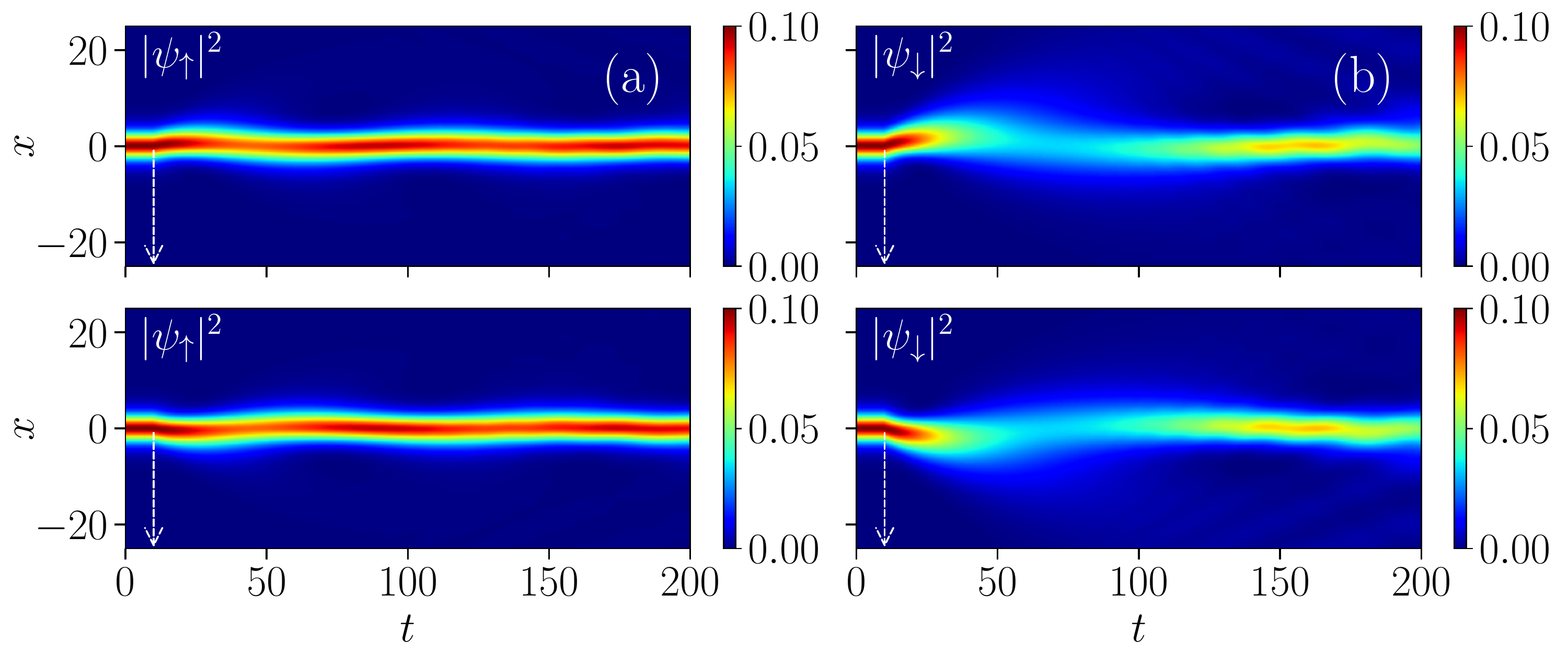}
\end{center}
\caption{Time evolution of the bright soliton of $\vert \psi_{\uparrow}\vert ^2$ (top panel) and $\vert \psi_{\downarrow}\vert ^2$ (bottom panel) due to sudden introduction of spin-orbit coupling as (a) $k_L = 0 \to 0.1$, and (b) $k_L = 0 \to 0.2$ with the absence of Rabi coupling ($\Omega = 0$). The dashed vertical line marks the time instant ($t = 10$) at which the quench is performed. The initial wavefunctions are prepared with $\alpha = \beta = -0.8$, $k_L = 0$, $\Omega = 0$, $\lambda = 0$, and correspond to $\mu_{\uparrow} = \mu_{\downarrow} \approx -0.082$.}
\label{fig5}
\end{figure}

Figures~\ref{fig5}(a) and \ref{fig5}(b) illustrate the dynamics of the soliton of the spin components, $\vert \psi_{\uparrow}\vert ^2$ (top panel) and $\vert \psi_{\downarrow}\vert ^2$ (bottom panel), due to the change in the spin-orbit coupling from $k_L=0$ to (a) $0.1$ and (b) $0.2$, respectively, at time $t=10$, while keeping Rabi coupling term as zero ($\Omega = 0$). During time evolution, the spin densities execute periodic oscillations both in space and time while preserving the symmetry property of the spin components, that is, $\psi_\uparrow(x) = \psi_\downarrow(-x)$. However, for the case of larger $k_L$, for example $k_L \geq 0.5$, the spin densities tend to propagate in the opposite directions. That is, $\vert \psi_{\uparrow}\vert ^2$ moves in the $+x$ direction while $\vert \psi_{\downarrow}\vert ^2$ propagates to $-x$ direction. %

The sudden introduction of spin-orbit coupling adds opposite phase factors, of the form $\mathrm{e}^{\pm i k_L x}$, in the spin components, which in turn contributes to the group velocity and thereby leads to the motion of solitons. This dephasing steers the decay of solitons as time progresses, while a revival occurs due to spin flipping. From the semiconductor physics point of view, the caveats in spin-based applications for quantum information processing are spin-relaxation and spin dephasing, which is the decay or loss of coherence of spin components~\cite{MWWu2010}. Hence, it will be of interest to explore the decay and revival of spin densities. One may note that a previous study reports the decay of solitons by evolving the exact soliton solution for a shorter time duration with the introduction of spin-orbit coupling~\cite{Wen2016}. However, here we witness the decay as well as the revival of soliton over longer times as a result of spin-orbit coupling. Further, in the absence of SO coupling, the spin densities propagate with a group velocity independent of their spin. This phenomenon could be useful for performing spin-based logical operations.

We also study the dynamics by simultaneously quenching both the SO and Rabi coupling strengths. %
\begin{figure}[!ht]
\begin{center}
\includegraphics[width=0.99\linewidth]{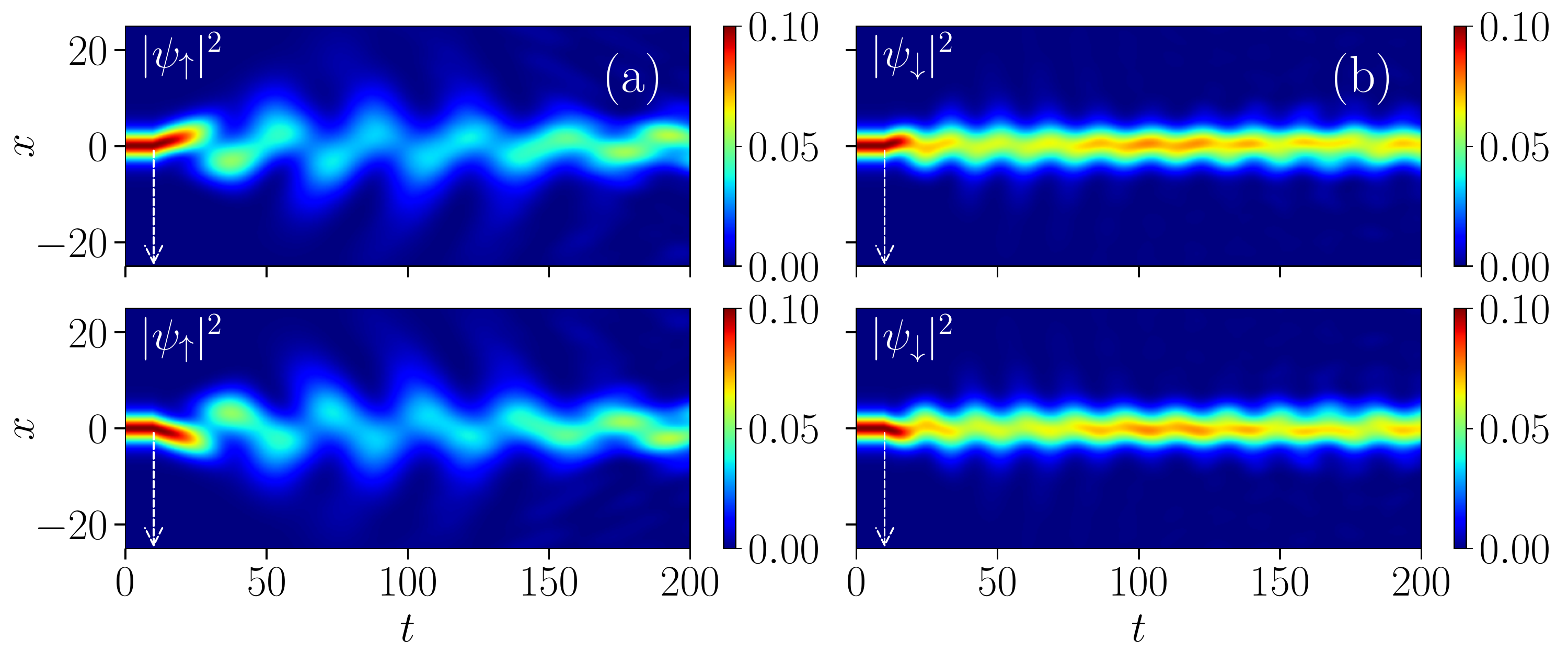}
\end{center}
\caption{Time evolution of the spin densities $\vert \psi_{\uparrow}\vert ^2$ (top panel), and $\vert \psi_{\downarrow}\vert ^2$ (bottom panel) illustrating decay and revival due to the effect of introducing both SO and Rabi coupling simultaneously as (a) $k_L = 0 \to 0.2$ and $\Omega = 0 \to 0.1$ and (b) $k_L = 0 \to 0.2$ and $\Omega = 0 \to 0.2$ at $t=10$. The system is prepared in its ground state with set of parameters as given in figure \ref{fig5}.}
\label{fig6}
\end{figure}%
We notice that the soliton dynamics is strongly dependent on the choice of $k_L$ and $\Omega$. For small values of $\Omega$, the densities show stable oscillations, while decay and revival of solitons are witnessed for dominant spin-orbit effect. In figure \ref{fig6}, we illustrate the dynamics by switching on both $k_L$ and $\Omega$ together at $t = 10$. Figure \ref{fig6}(a) shows the evolution of the spin densities by instantly changing $k_L$ from $0 \to 0.2$ and $\Omega$ from $ 0 \to 0.1$. Here, the soliton initially moves due to $k_L$ and, as time progresses, it starts to oscillate both in space and time, as a result of Rabi coupling. Also, there is a sequence of a decay followed by revival in the densities and the time interval of successive decay and revival remains almost constant. In Figure \ref{fig6}(b), we show the dynamics for the case with $k_L \to 0.2$ and $\Omega \to 0.2$. However, this small increase in $\Omega$ leads the solitons to execute much smaller spatial oscillations and the frequency of successive decay and revival process gets doubled. Further, during the time evolution, the spin-densities preserve the cross symmetry, that is, $\vert \psi_{\uparrow} (x) \vert ^2 = \vert \psi_{\downarrow}(-x)\vert ^2$.  The role of both SO and Rabi coupling parameters is to produce either decay and revival of solitons for $\Omega > k_L^2$ or collapse when $\Omega < k_L^2$.

\begin{figure}[!ht]
\begin{center}
\includegraphics[width=0.99\linewidth]{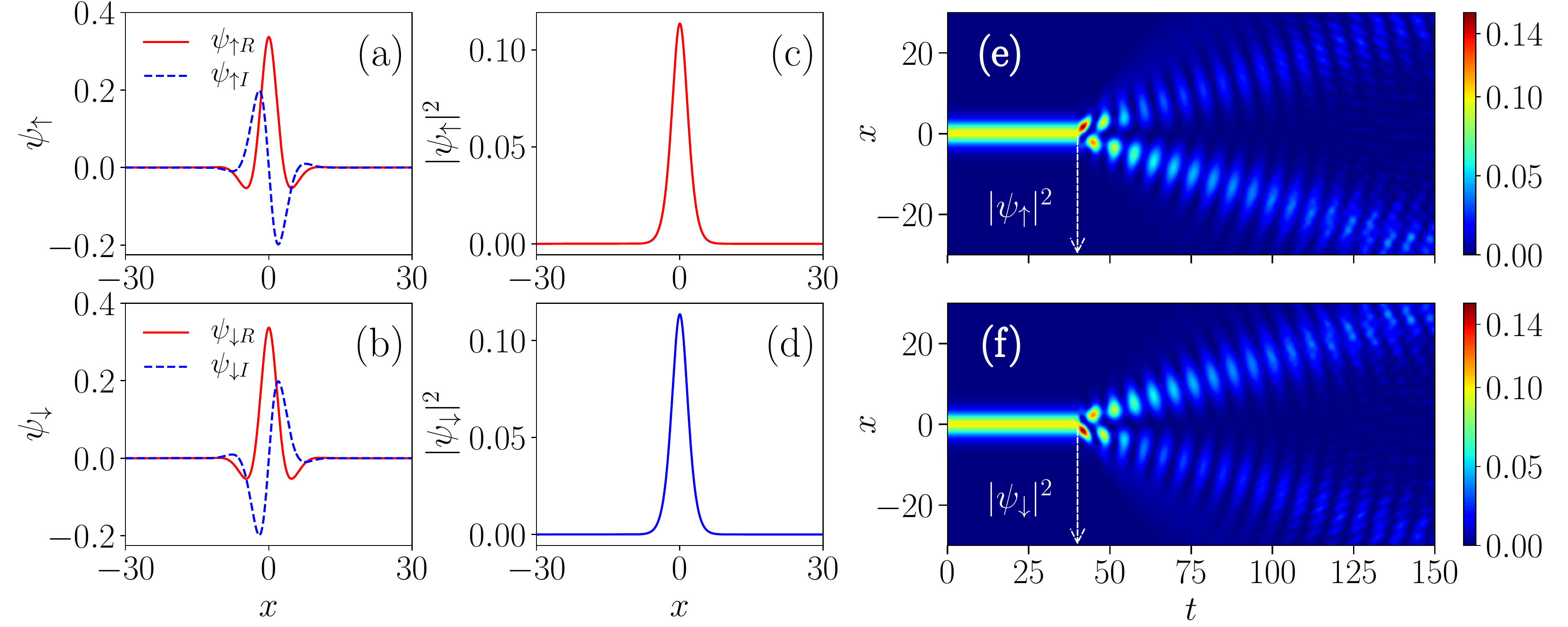}
\end{center}
\caption{Plots of the real (solid red line) and the imaginary (dashed blue line) parts of the initial wavefunctions, (a) $\psi_{\uparrow}$ and (b) $\psi_{\downarrow}$ and the spin densities (c) $\vert \psi_{\uparrow} \vert^2$ and (d) $\vert \psi_{\downarrow} \vert^2$ prepared with $\alpha = \beta = -0.8$, $k_L = 0.5$, $\Omega = 0$, and $\mu_{\uparrow} = \mu_{\downarrow} \approx 0.205$. (e) - (f): Spatiotemporal dynamics showing the instability due to the quench by increasing the Rabi coupling strength $\Omega$ from $0$ to $0.5$ at $t=40$.}
\label{fig7}
\end{figure}

Next, we investigate the effect of increasing the Rabi coupling strength while keeping the spin-orbit coupling constant. For this purpose, we prepare a profile of stationary wavefunctions by fixing $k_L = 0.5$ and $\Omega = 0$ with with $\alpha = \beta = -0.8$, and $\lambda = 0$. We then evolve this stationary profile and the system is quenched by suddenly applying the Rabi coupling at time $t = 40$. Figures~\ref{fig7}(a) and \ref{fig7}(b) show the real and imaginary parts of the initial wavefunctions, and  figures~\ref{fig7}(c) and \ref{fig7}(d) depict the corresponding density profiles of the stationary bright solitons. Figures~\ref{fig7}(e) and \ref{fig7}(f) illustrate the time evolution of the bright soliton of the spin-up and spin-down states, respectively. %
As the initial profile is prepared with finite $k_L$, the spin wavefunctions contain both real and imaginary parts, which lead the solitons to propagate in the respective directions due to spin orbit coupling. However, the presence of Rabi coupling creates an oscillatory instability~\cite{Jiang2016}, which converts the density into breather-like pulses as shown in  figures~\ref{fig7}(e) and \ref{fig7}(f). 

In all the above cases, the decay of soliton is due to the violation of Galilean invariance by the spin-orbit coupling~\cite{Xu2013,Zhu2012}. We have demonstrated how the decay can be controlled suitably by quenching the spin-orbit and Rabi coupling strengths. 

\subsection{Quenching dynamics with repulsive inter- and intra component interactions}
\label{sec:5:2}
In the previous section, we have studied soliton dynamics by the abrupt shift in the coupling strengths for the case of attractive intra- and inter-component interactions. Next, we shall examine the case of spin-orbit coupled BECs with repulsive intra- and inter-component interactions. It may be noted that, due to the repulsive nature of the interactions, the spin components tend to expand in the absence of trap. The inclusion of SO coupling makes this expansion faster, and hence it becomes necessary to apply a weak harmonic trap potential to stabilize the expanding condensate. With repulsive interactions, the SO coupled BECs exhibit three distinct phases, namely plane wave, zero momentum, and stripe wave phases, as identified from the single-particle spectrum~\cite{Achilleos2013-ds, YLi2012}.

\begin{figure}[!ht]
\begin{center}
\includegraphics[width=0.99\linewidth]{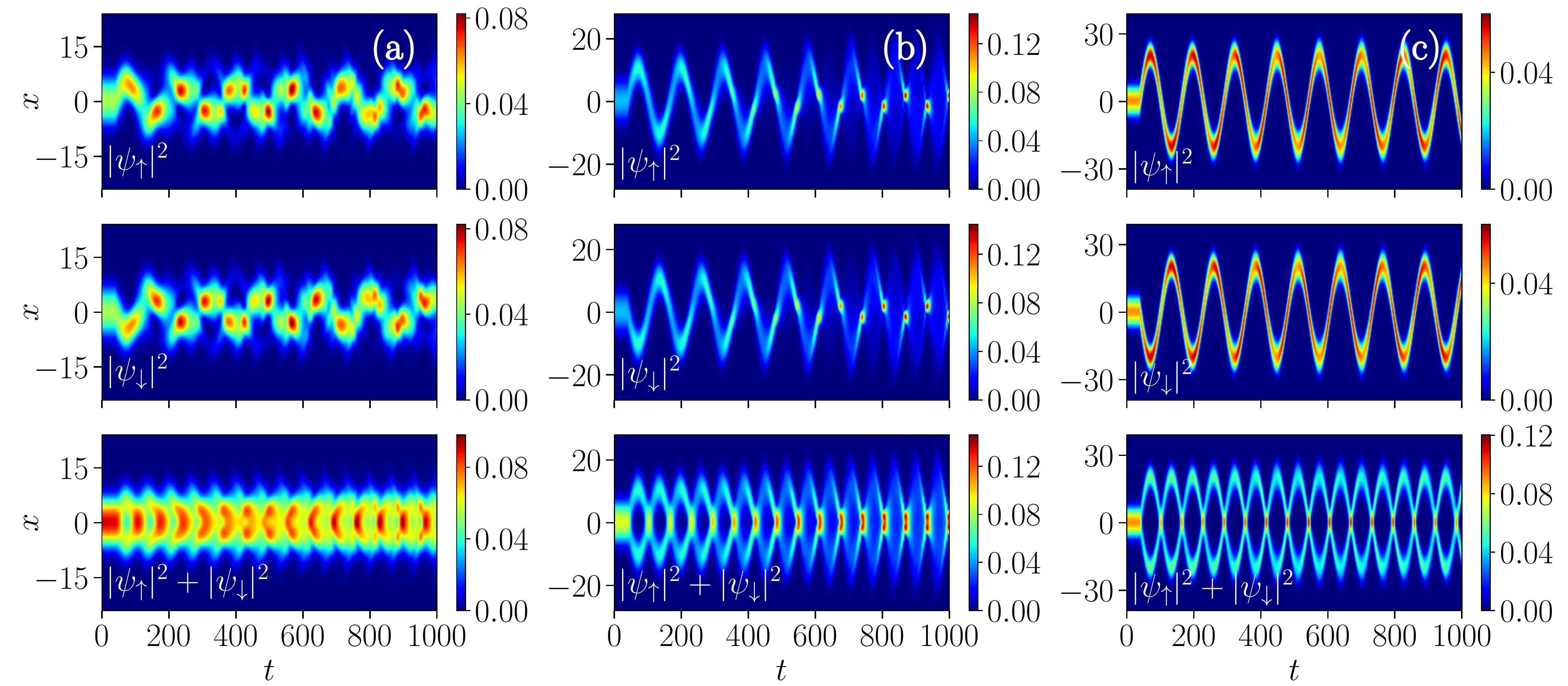}
\end{center}
\caption{Spatio-temporal dynamics of the densities of the spin components $\vert \psi_{\uparrow}\vert ^2$ (top row), $\vert \psi_{\downarrow}\vert ^2$ (middle row), and the total density $\vert \psi_{\uparrow}\vert ^2 + \vert \psi_{\downarrow}\vert ^2$ (bottom row) illustrating the dynamical transition from spin-mixed to spin separated states. The stationary state is prepared with $\alpha = \beta = 0.8$, $\lambda = 0.05$, $k_L = 0$, and $\Omega = 0$ with $\mu_{\uparrow} = \mu_{\downarrow} = 0.08$. The quenching is applied at $t=50$ by suddenly introducing SO coupling as (a) $k_L = 0 \to 0.2$, (b) $k_L = 0 \to 0.5$ and (c) $k_L = 0 \to 1.0$ at $t = 50$.}
\label{fig08}
\end{figure}

We prepare a stationary state wavefunction without SO and Rabi coupling, as discussed above, and evolve this stationary state to analyze the dynamics by switching on the SO and Rabi coupling strengths at a finite time. For instance, we numerically generate the time-independent wavefunctions (stationary profile) by fixing $\alpha = \beta = 0.8$, $\lambda = 0.05$, $k_L = 0$, and $\Omega = 0$. In Figure \ref{fig08}, we show the sptaiotemporal dynamics of the spin densities by introducing SO coupling of different strengths at $t=50$. For instance, when we introduce the SO coupling of strength $k_L = 0.2$ at $t=50$ %
the system transits from spin-mixed state to spin separated state and the spin densities execute broken oscillations as shown in figure \ref{fig08}(a). Both the spin-components (top and middle row) exhibit similar oscillatory motions and maintain the symmetry. Interestingly, the total density of the two spin components displays a standing wave pattern with a larger fraction of atoms oscillating near $x = 0$ as depicted in the bottom row of figure \ref{fig08}(a).

However, changing $k_L$ with slightly higher values, say for example $k_L = 0.5$, the localized spin densities near $x=0$ begin to oscillate within the trap as illustrated figure \ref{fig08}(b). Initially, in these condensed atoms, both the spin-components exhibit oscillations, and as time progress they accumulate while approaching the trap center. When compared to the case of quenching at $k_L = 0.2$, after a sufficiently longer time the total density profile shows prominent standing wave pattern, with accumulated density near the trap center, as shown in the bottom row of figure \ref{fig08}(b).

When quenching the system with even larger $k_L$ values, the oscillation of the condensate becomes more regular, and the two spin components exhibit oscillatory wave patterns as shown in figure \ref{fig08}(c). However, the total density has a localized maximum at the nodes as shown in figure \ref{fig08}(c) [bottom row], which implies that the spin components are fully mixed at the center of the trap. Whereas at the anti-nodes, the maximum density is equally shared between the components revealing the spin-separated state. Thus, the system exhibits periodic spin mixing-demixing dynamics when quenching with large $k_L$ values.  Similar periodic oscillations with relatively smaller amplitudes have been observed when changing both $k_L$ and $\Omega$ simultaneously.

We also investigate the quench dynamics of stationary state prepared with SO coupling for the case of repulsive atomic interactions. %
\begin{figure}[!ht]
\begin{center}
\includegraphics[width=0.99\linewidth]{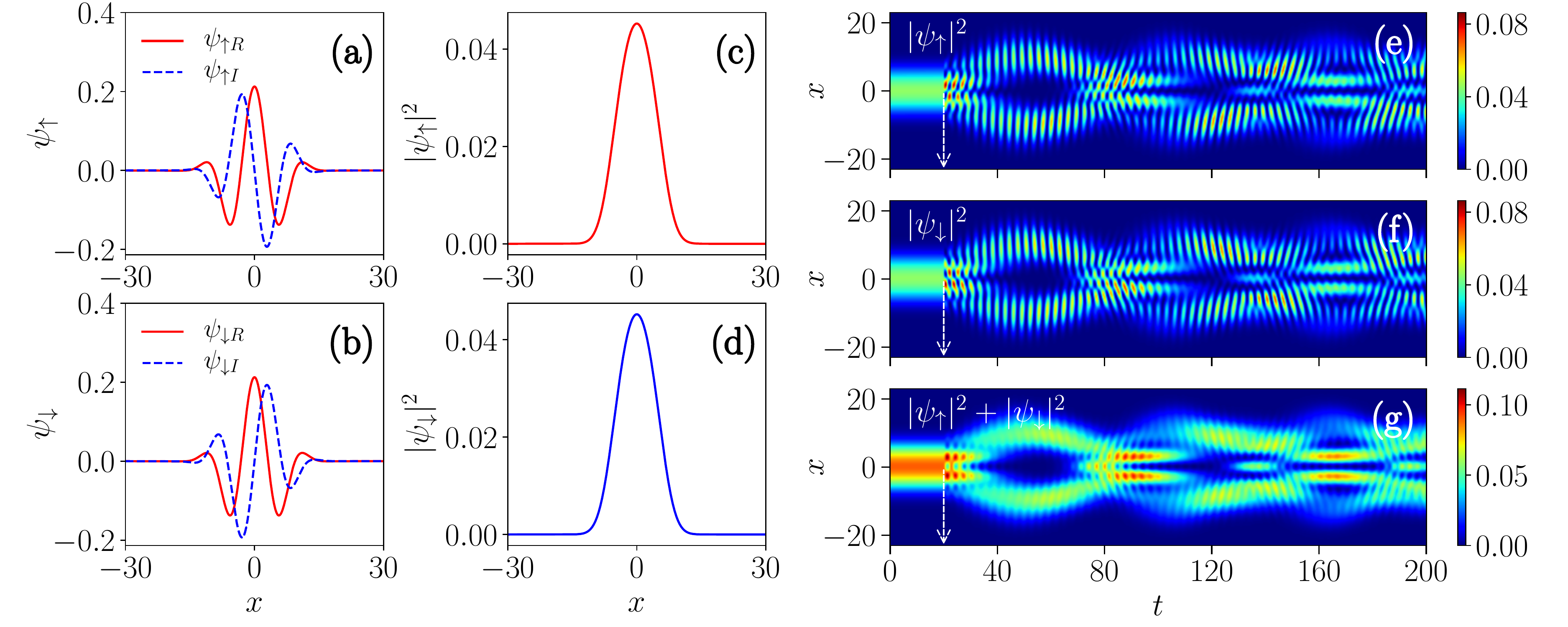}
\end{center}
\caption{Plots of the real (solid red line) and the imaginary (dashed blue line) parts of the stationary wavefunctions, (a) $\psi_{\uparrow}$, (b) $\psi_{\downarrow}$, and the densities (c) $\vert \psi_{\uparrow}\vert ^2$ and (d) $\vert \psi_{\downarrow}\vert ^2$ for $\lambda = 0.05$, $\alpha = \beta = 0.8$, $k_L = 0.5$, and $\Omega = 0$ with $\mu_{\uparrow} = \mu_{\downarrow} \approx -0.0427$. Plots (e) and (f) show the time evolution of the spin densities and (g) is the total density $\vert \psi_{\uparrow}\vert ^2 + \vert \psi_{\downarrow}\vert ^2$. Here the Rabi coupling with strength $\Omega = 1.0$ is turned on at $t = 20$.}
\label{fig09}
\end{figure}%
Figures~\ref{fig09}(a) - \ref{fig09}(d) depict the real and the imaginary parts of the stationary wavefunctions, and the density profiles obtained for $k_L = 0.5, \Omega = 0$, $\alpha = \beta = 0.8$, $\lambda = 0.05$, and $\mu_{\uparrow} = \mu_{\downarrow} = -0.0427$. We evolve this stationary state in real-time propagation by introducing the Rabi coupling $\Omega = 1.0$ at time $t = 20$. We observe that the homogeneous density cloud inside the trap splits into two halves about $x = 0$ after the quenching is applied. The spatially segregated atomic clouds elongate and move along the positive and negative $x$-directions. When the Rabi coupling is switched on, the condensate starts to oscillate both in space and time and exhibits a breathing-like motion. Initially, each of the spin-component splits into two parts, then they approach each other, interact and split again. These parts travel for a short time and execute temporal oscillations in the densities. After a finite interval of time, these clouds approach each other and intersect at $x = 0$ then they split into four density clouds with two located near to the trap center and the other two are in the lateral position. The clouds close to the trap center initially have the maximum density for a short time then they mix with the other clouds as time progress. This scenario repeats at a regular interval of time as shown in  figures~\ref{fig09}(e) and \ref{fig09}(f). 

\begin{figure}[!ht]
\begin{center}
\includegraphics[width=0.99\linewidth]{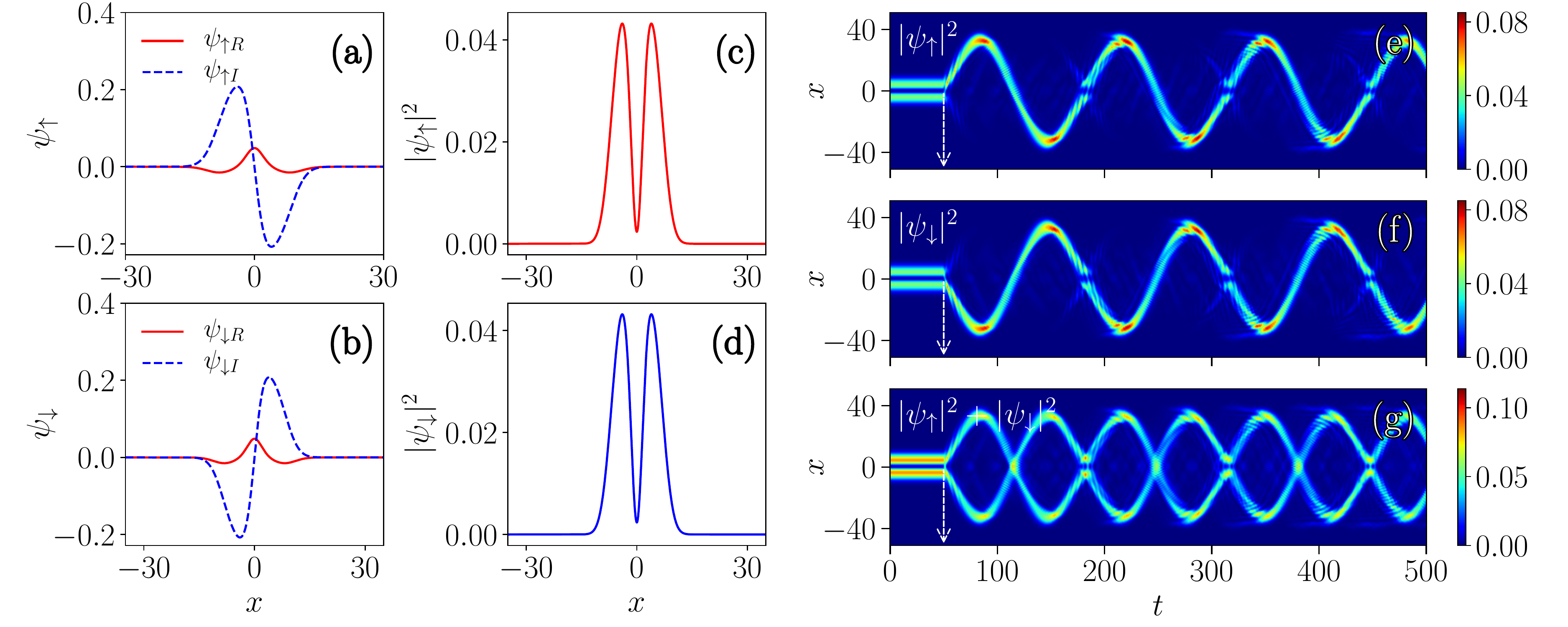}
\end{center}
\caption{Plots showing the real (solid red line) and the imaginary (dashed blue line) parts of the initial wavefunctions, (a) $\psi_{\uparrow}$ and (b) $\psi_{\downarrow}$ and the corresponding densities (c) $\vert \psi_{\uparrow}\vert ^2$, (d) $\vert \psi_{\downarrow}\vert ^2$ for $\lambda = 0.05$, $\alpha = \beta = 0.8$, $k_L = 0.5$, and $\Omega = 0.5$ with $\mu_{\uparrow} = \mu_{\downarrow} \approx -0.393$. (e) - (g) depict the time evolution of the densities of the spin components and the total density. Here the SO coupling strength ($k_L$) is suddenly increased from $0.5$ to $2.0$ at $t = 50$.}
\label{fig10}
\end{figure}

Also, we study the quench dynamics of the stationary state with $k_L = 0.5$ and $\Omega = 0.5$. The real and the imaginary parts of the stationary profile are shown in  figures~\ref{fig10}(a) and \ref{fig10}(b), and the corresponding spin densities are plotted in  figures~\ref{fig10}(c) and \ref{fig10}(d). %
Upon increasing the value of $k_L$ from $0.5$ to $2.0$ suddenly at $t=50$, either one of the two maxima of this spin-mixed state in each spin component fades away and the other starts to oscillate smoothly as illustrated in  figures~\ref{fig10}(e) and \ref{fig10}(f). Here also the total density exhibits a standing wave pattern as shown in figure \ref{fig10}(g).  For $k_L < 2$, it exhibits a similar dynamics but the density profiles oscillate in an irregular manner. This result is quite similar to that reported by Li et al.~\cite{CHLi2019}. 

\begin{figure}[!ht]
\begin{center}
\includegraphics[width=0.99\linewidth]{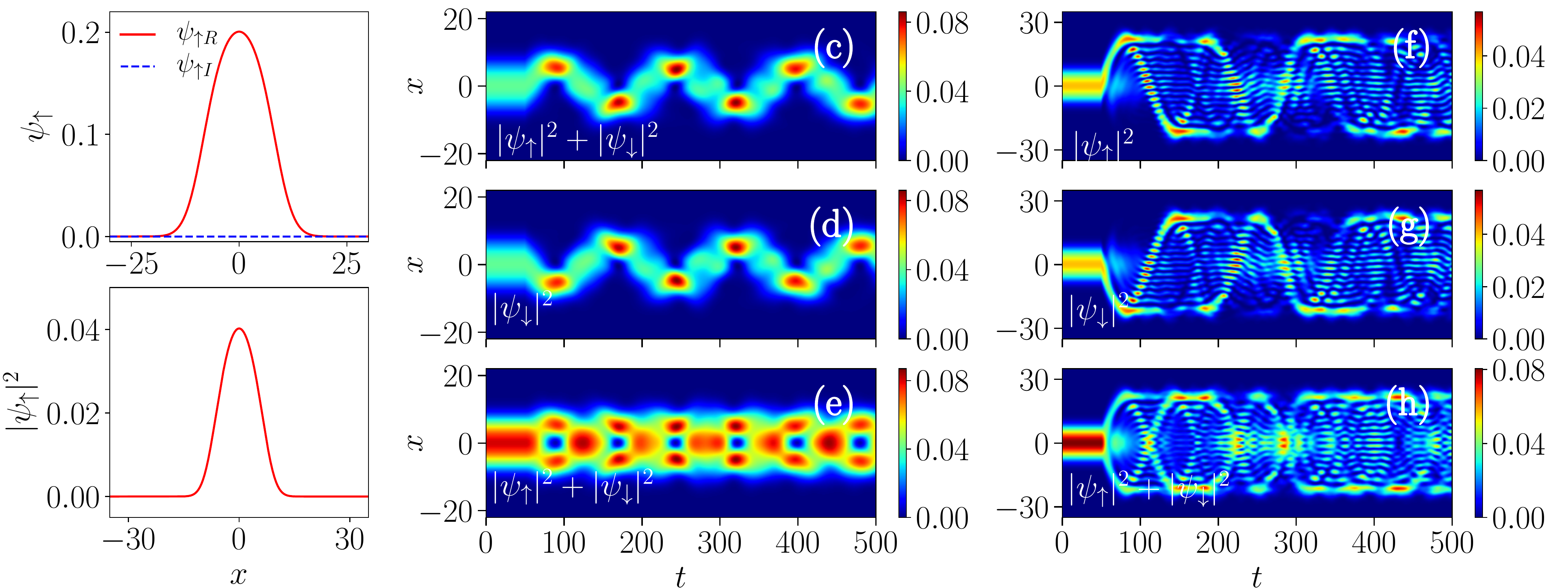}
\end{center}
\caption{Plots showing the real (solid red line) and the imaginary (dashed blue line) parts of the initial wavefunction, (a) $\psi_{\uparrow}$ and the density profile in (b) $\vert \psi_{\uparrow}\vert ^2$ for the the spin-up component for $\lambda = 0.05$, $\alpha =0.5, \beta = 2.0$, $k_L = 0$, and $\Omega = 0$ with $\mu_{\uparrow} = \mu_{\downarrow} \approx 0.108$. Spin densities (c) $\vert \psi_{\uparrow}\vert ^2$, (d) $\vert \psi_{\downarrow}\vert ^2$ and (e) total density $\vert \psi_{\uparrow}\vert ^2 + \vert \psi_{\downarrow}\vert ^2$ exhibiting periodic oscillations due to the introduction of SO coupling $k_L = 0.2$ at $t=50$. Plots (f) - (h) illustrate the dynamics of spin densities due to the sudden introduction of $k_L = 1.0$ and $\Omega = 0.1$ at $t=50$.}
\label{fig11}
\end{figure}

Additionally, we consider the case of binary BECs with $\alpha = 0.5$ and $\beta = 2.0$, and prepared the stationary wavefunction for $k_L = 0$ and $\Omega = 0$. Figures~\ref{fig11}(a) and \ref{fig11}(b) depict the real and imaginary parts of the stationary wavefunction, $\psi_\uparrow$ and the corresponding density, $\vert\psi_\uparrow \vert^2$. In this case, the wavefunctions of the spin components are equal, that is $\psi_\uparrow = \psi_\downarrow$. When the system is quenched at $k_L = 0.2$, %
the spin-densities show periodic oscillations with maximum density at the extrema as illustrated in  figures~\ref{fig11}(c) and \ref{fig11}(d). From the total density shown in figure \ref{fig11}(e), we notice that once the Rabi coupling parameter is altered, the spin-mixed state exhibits a sequence of periodic patterns. %
On the other hand, setting $k_L =1.0$ with $\Omega = 0$ for $\alpha = 0.5$, $\beta = 2.0$, the system exhibits similar dynamics as of the case with $\alpha = 0.8, \beta = 0.8$ (see figure \ref{fig08} above). However, the frequency of oscillation is relatively small compared to that shown in figure \ref{fig08}(c).

Next, we investigate the dynamics by applying Rabi and SO coupling parameters concurrently. When $\Omega = 0 \to 0.01$, and $k_L = 0 \to 1.0$ the spin-components initially starts to oscillate and crossover each other without interaction. After sufficiently longer time, they interfere with each other and form a stripe pattern in the density. Further, in the system quenched by a slightly higher value of Rabi coupling, say for example $\Omega = 0 \to 0.1$, the spin densities separate for a while, then oscillate and exhibit a sequence of interference patterns as shown in  figures~\ref{fig11}(f) - \ref{fig11}(h). 
Similar interference pattern formation has been observed and reported in spin-orbit coupled BECs~\cite{CHLi2019}. This kind of interference pattern stems due to Rabi coupling and can be suppressed either by decreasing $k_L$ or by increasing $\Omega$.

So far we have considered weak or moderate interaction strengths ($\alpha$ and $\beta$) and studied the dynamics. However, it is natural to extend the analysis of quenching dynamics for strong interactions. In the following, we investigate the dynamics by suddenly altering the coupling parameters with strong repulsive interactions. For instance, we prepare a stationary profile of a binary BEC ($k_L =0$ and $\Omega = 0$) with the interaction strengths $\alpha = 5$, $\beta = 25.0$, and with a weak trap ($\lambda = 0.05$). %
\begin{figure}[!ht]
\begin{center}
\includegraphics[width=0.99\linewidth]{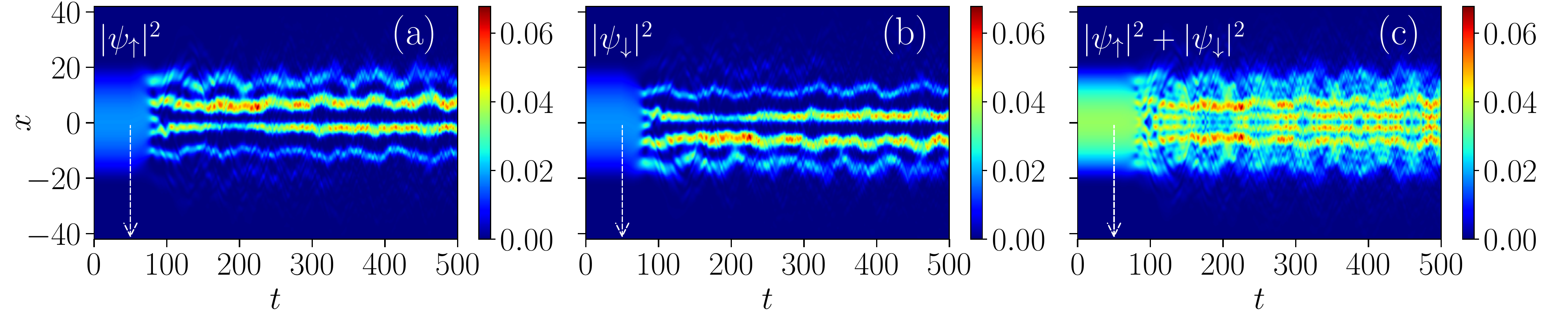}
\end{center}
\caption{Plots showing the filament formation: spin densities (a) $\vert \psi_{\uparrow}\vert ^2$, (b) $\vert \psi_{\downarrow}\vert ^2$, and the total density (c) $\vert \psi_{\uparrow}\vert ^2 + \vert \psi_{\downarrow}\vert ^2$. The system is prepared in its ground state with set of parameters as $\lambda = 0.05$, $\alpha = 5, \beta = 25$, $k_L = 0$, and $\Omega = 0$ with $\mu_{\uparrow} = \mu_{\downarrow} \approx 0.542$. Here the SO coupling with strength $k_L = 0.1$ is introduced suddenly at $t = 50$.}
\label{fig12}
\end{figure}%
In figure~\ref{fig12}, we show the time evolution of spin densities by switching on the spin-orbit coupling, i.e. $k_L = 0 \to 0.1$, while keeping a zero Rabi coupling strength ($\Omega = 0$) at $t=50$. In this case, the spin densities propagate for a short duration and segregate, as time progresses this leads to filament formation, these filaments are in an immiscible state or a dark-bright state~\cite{Mistakidis2018, HKiehn2019}. In other words, we observed a miscible-to-immiscible phase transition. A similar kind of filament formation in binary BECs reported in the literature~\cite{Navarro2009}. This filament formation is due to the strength of inter-species interaction, which is larger than that of intra-species interaction.

However, quenching the system with higher values of spin-orbit coupling strength results in a reduced number of filaments. Similar behaviour in the dynamics observed when introducing the Rabi coupling. It shows filament formation for $\Omega < k_L^2$, while a different nonlinear pattern appears when $\Omega > k_L^2$. We also noticed that the time of quench alters the number of filaments. For instance, if we introduce the SO coupling at $t=80$ instead of $t=50$, the number of filaments gets increased. Besides, these dynamics are observed in the weak coupling limit only, while the strong repulsion in a weak trap destabilizes the condensate.

\section{Dynamics of spin-orbit coupled BEC\lowercase{s} by quenching spin-orbit and Rabi coupling parameters}
\label{sec:6}
In the above, we have studied the dynamics of both binary and spin-orbit coupled BECs by sudden introduction of spin-orbit and (or) or Rabi coupling parameters. However, it will be worth to investigate the dynamics of spin-orbit coupled BECs by an instantaneous change in the either SO or Rabi coupling parameters. A stationary profile of condensate wavefunctions is prepared with non-zero SO and Rabi coupling parameters by the imaginary time propagation. Then we study the time evolution of this profile by changing these parameters. In this case, one can think of two different types of quenching by which the parameter is changed.

\begin{figure}[!ht]
\begin{center}
\includegraphics[width=0.99\linewidth]{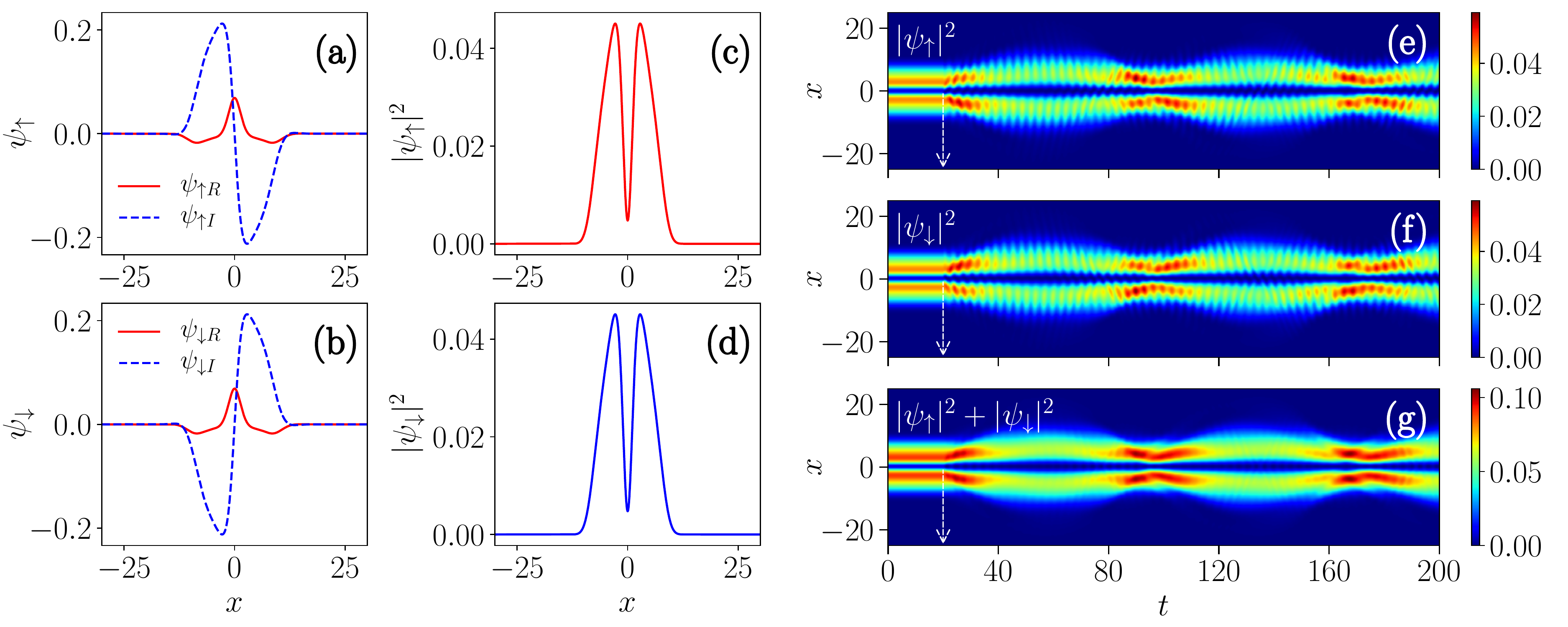}
\end{center}
\caption{Plots of the real (solid red line) and the imaginary (dashed blue line) parts of the plane wavefunctions, (a) $\psi_{\uparrow}$ and (b) $\psi_{\downarrow}$ and the corresponding densities (c) $\vert \psi_{\uparrow}\vert ^2$ and (d) $\vert \psi_{\downarrow}\vert ^2$ for $\lambda = 0.05$, $\alpha =0.5, \beta = 2.0$, $k_L = 1.0$, and $\Omega = 1.0$ with $\mu_{\uparrow} = \mu_{\downarrow} \approx -0.886 $. Time evolution of (e) $\vert \psi_{\uparrow}\vert ^2$, (f) $\vert \psi_{\downarrow}\vert ^2$ and (g) total density $\vert \psi_{\uparrow}\vert ^2 +\vert \psi_{\downarrow}\vert ^2$ showing a breather-like oscillations. Here the SO coupling strength decreased from $k_L = 1.0$ to $0.5$ at $t = 20$.}
\label{fig13}%
\end{figure}

In the first type, we initially prepare a stationary profile of the SO coupled BECs with fixed $k_L$ and $\Omega$, and during time evolution the parameter ($k_L$ or $\Omega$) is decreased. Whereas, in the second type, is done by increasing the value of either $k_L$ or $\Omega$. In  figures~\ref{fig13}(a) - \ref{fig13}(d), we show the typical stationary plane wave profiles of the two spin components and their densities for the choice of parameters $\alpha =0.5$, $\beta =2$, $\lambda = 0.05$, $\Omega = 1$, and $k_L=1$. We evolve these profiles by either decreasing or increasing the coupling parameters. For instance, we lower the value of $k_L$ from $1$ to $0.5$ while keeping the remaining parameters unaltered.

During this process the density profiles of both the spin components exhibit breather like oscillations with a dip in the density at the center of the trap, as shown in  figures~\ref{fig13}(e) - \ref{fig13}(g).  We also drive the system by the rapid increase of $k_L$ from $1$ to $1.5$ while keeping $\Omega = 1$. %
In this case, the density profile displays similar patterns reported by Li et al.~\cite{CHLi2019}.  Further, the spin-orbit coupled BECs exhibits breather like oscillations when we increase the value of $\Omega$ from $1$ to $1.5$ while keeping the other parameters the same as before. We have also observed certain nonlinear wave patterns when the system is quenched by suddenly decreasing $\Omega$ from $1$ to $0.5$. Based on the observations, we conclude that the system shows breather like oscillating density profile for $\Omega > k_{L}^{2}$, and nonlinear wave patterns for $\Omega < k_{L}^{2}$.

Next, we prepare a stationary profile of wavefunctions for larger values of the interactions strengths, for example, $\alpha = 5$ and $\beta = 25$, and the coupling parameters are kept as $\Omega=1$ and $k_L = 1$. One may note that new higher order excited states are possible especially, for weak trap with intra- and inter-species interaction strengths obeying the condition $\beta > \alpha$~\cite{Navarro2009}. These higher order states are generally spin-mixed states. The emergence of these higher order excited states is due to the manifestation of an effective modulational instability. For larger interspecies interaction strengths, higher modulational wave numbers become unstable. 

\begin{figure}[!ht]
\begin{center}
\includegraphics[width=0.99\linewidth]{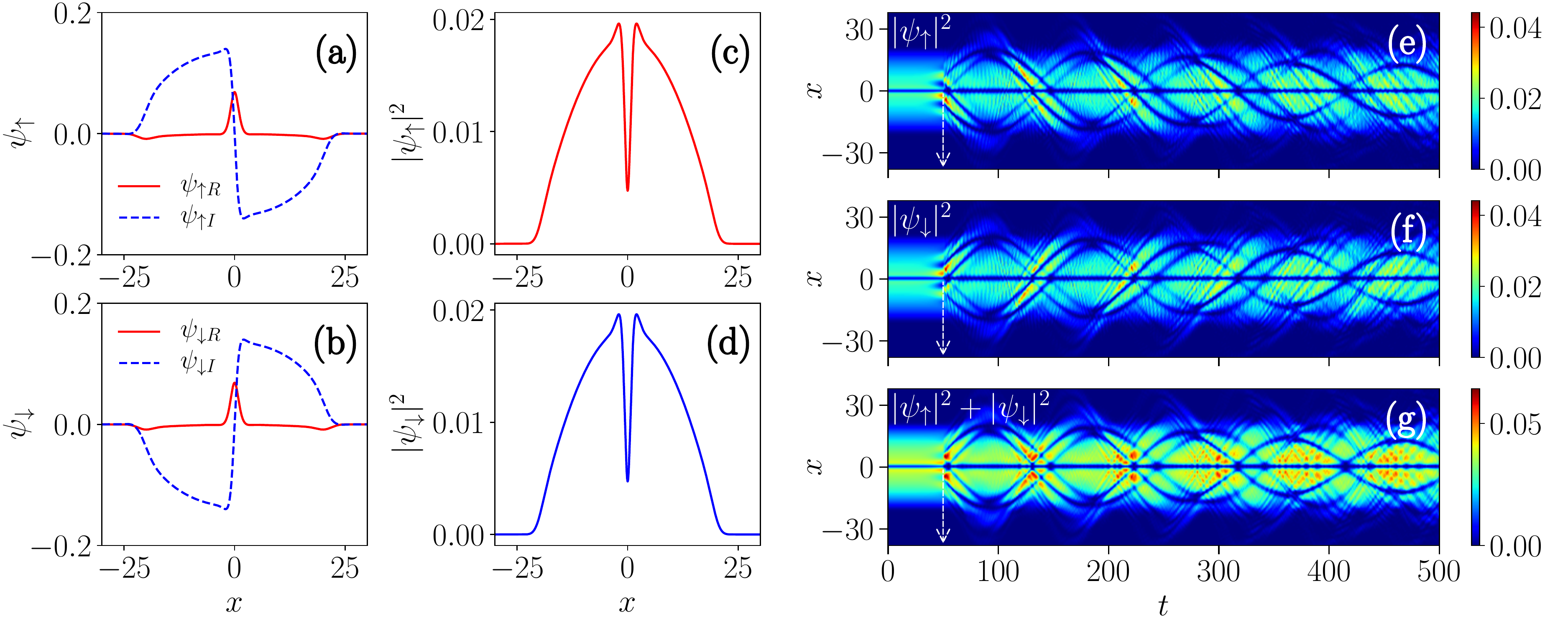}
\end{center}
\caption{Plots of the real (solid red line) and the imaginary (dashed blue line) parts of the plane wave wavefunctions for $\lambda = 0.05$, $\alpha = 5$, $\beta = 25$, $k_L = 1.0$, and $\Omega = 1.0$ with $\mu_{\uparrow} = \mu_{\downarrow} \approx -0.448$: (a) $\psi_{\uparrow}$ and (b) $\psi_{\downarrow}$, (c) $\vert \psi_{\uparrow}\vert ^2$ and (d) $\vert \psi_{\downarrow}\vert ^2$. Time evolution of the density profiles of the spin components (e) $\vert \psi_{\uparrow}\vert ^2$ and (f) $\vert \psi_{\downarrow}\vert ^2$, and total density (g) $\vert \psi_{\uparrow}\vert ^2 + \vert \psi_{\downarrow}\vert ^2$ illustrating quench induced dark soliton dynamics. Here the quenching is done by decreasing $k_L$ from $ 1.0$ to $0.5$ at $t = 50$.}
\label{fig14}
\end{figure}

We generate a profile of wavefunctions through imaginary time propagation by fixing the parameters as $\lambda = 0.05$, $\alpha = 5$ and $\beta = 25$, $\Omega=1$ and $k_L = 1$. In  figures~\ref{fig14}(a) and \ref{fig14}(b), we show the real and imaginary parts of the wavefunctions of the spin components, and in  figures~\ref{fig14}(c) and \ref{fig14}(d) we plot the corresponding density profiles. %
It is easy to see from  figures~\ref{fig14}(a) and \ref{fig14}(b) that these wavefunctions obey cross-symmetry, that is, $\psi_{\uparrow}(x) = \psi_{\downarrow}(-x)$. Further, these profiles live for short durations and become unstable during time evolution due to strong repulsion. However, this can be stabilized by suitably quenching the coupling parameters. For instance, an abrupt change in the coupling parameters at the time of instability during time evolution results in the formation of stable dark solitons. Figures~\ref{fig14}(e) - \ref{fig14}(g) demonstrate the dynamically created dark solitons by reducing $k_L$ from $1$ to $0.5$ at $t=50$. In general, dark solitons are observed at the interface between the phase domains by engineering a phase difference in a condensate~\cite{Burger1999, Becker2008}, and multiple dark solitons are created by combining two coherent condensates~\cite{Weller2008, Theocharis2008}. Dark solitons can also be created by driving the system from equilibrium to nonequilibrium~\cite{Zurek2009, Damski2010}. We just demonstrate another possible way of creating dark solitons by lowering SO coupling strength. We also observe the formation of dark solitons as long as $\Omega > k_{L}^{2}$ is maintained and the quench is applied near the time of instability.

We have also verified the creation of dark solitons for a different set of an anti-symmetric initial profile. A similar dynamics with relatively lower number of dark solitons is observed with an anti-symmetric stationary profile, that is, $\psi_{\uparrow}(x) = -\psi_{\downarrow}(-x)$. However, for spatially separated density profiles, a large number of dark solitons can be created with both symmetric as well as anti-symmetric initial profiles.   

Furthermore, we extend the study of dynamics for the stripe phase case by preparing the stationary state wavefunctions with $\alpha = \beta = 0.8$ and the coupling strengths $k_{L} = 2$ and $\Omega = 1$. %
\begin{figure}[!ht]
\begin{center}
\includegraphics[width=0.99\linewidth]{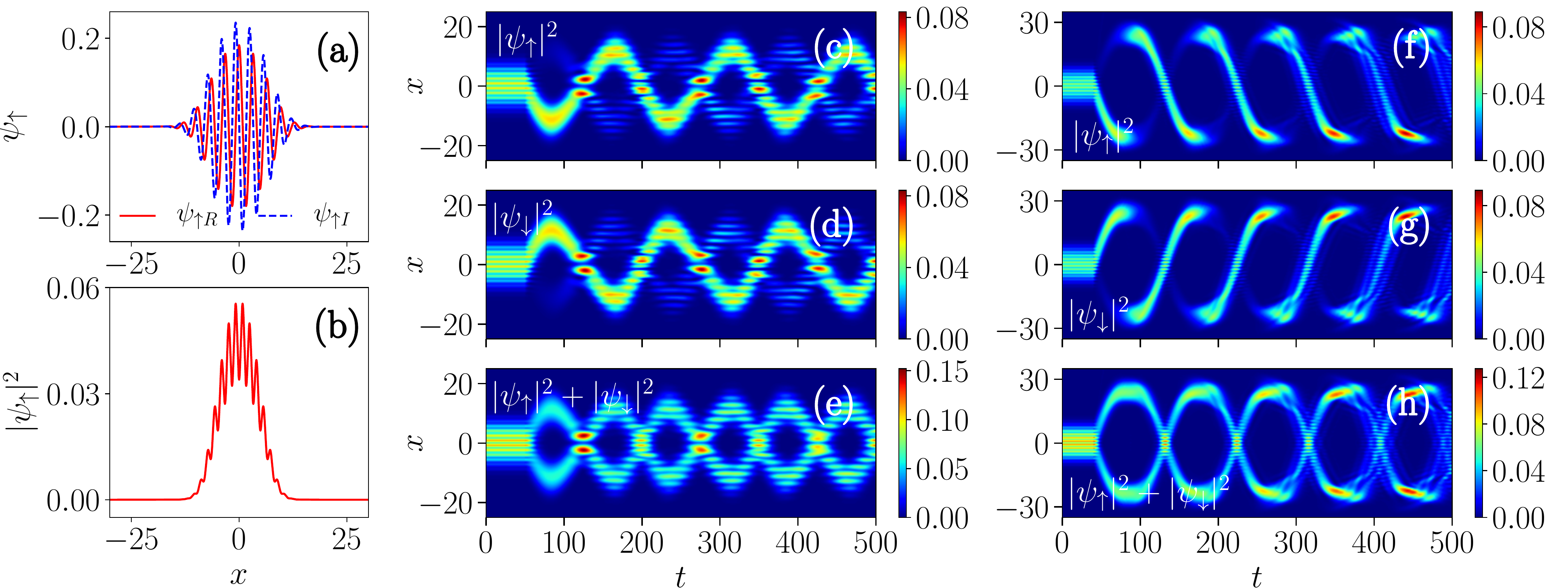}
\end{center}
\caption{Plots of real (solid red line) and the imaginary (dashed blue line) parts of the stripe wavefunctions for $\lambda = 0.05$, $\alpha = \beta = 0.8$, $k_L = 2.0$, and $\Omega = 1.0$ with $\mu_{\uparrow} = \mu_{\downarrow} \approx -2.041$: (a) wavefunction $\psi_{\uparrow}(x)$ and (b) spin density $\vert \psi_{\uparrow}(x)\vert ^2$ of the spin-up component. Time evolution of the spin densities (c) $\vert \psi_{\uparrow}\vert ^2$, (d) $\vert \psi_{\downarrow}\vert ^2$, and (e) total density $\vert \psi_{\uparrow}(x)\vert ^2 + \vert \psi_{\downarrow}\vert ^2$ showing oscillatory stripe patterns due to sudden decrease of $k_L$ from $2$ to $1.5$ at $t = 50$. The induced spin-mixing and spin-flipping dynamics of the spin densities (f) $\vert \psi_{\uparrow}\vert ^2$, (g) $\vert \psi_{\downarrow}\vert ^2$, and (h) total density $\vert \psi_{\uparrow}\vert ^2 + \vert \psi_{\downarrow}\vert ^2$ due to sudden decrease of $k_L$ from $2$ to $1$ at t = 50.}
\label{fig15}
\end{figure}%
In  figures~\ref{fig15}(a) and \ref{fig15}(b), we show the real and imaginary parts of the stationary wavefunction and the density, respectively of the spin-up component. Note that the wavefunctions of the spin components obey anti-symmetry. Quenching the system by decreasing the value of $k_{L}$ from $2$ to $1.5$ results in the oscillatory stripe pattern in both the components as illustrated in
 figures~\ref{fig15}(c) - \ref{fig15}(e).

Finally, we consider the stripe phase case with different interaction strengths, say for example, $\alpha = 0.5$ and $\beta = 2$. The stationary profile is similar to that shown in figure \ref{fig15}(a). The system dynamics is modified by lowering the value of spin-orbit coupling strength $k_L$ from $2$ to $1$ which results in the stripe phase spin-mixed state turning into an oscillating spin separated state of equal densities. During the oscillation, the spin densities show spin flipping at the maxima and striped spin-mixing at the minima. As time progress, the stripe pattern at the minima expands as shown in  figures~\ref{fig15}(f) - \ref{fig15}(h). When lowering $k_{L}$ from $2$ to $1.5$, the density profiles exhibit nonlinear wave patterns. Further, raising $k_{L}$ from $2$ to $2.5$ the density profiles show oscillations with stripe wave modulation. For Rabi quenching, the system exhibits similar dynamics as shown in  figures~\ref{fig11}(c) - \ref{fig11}(e) with stripe wave modulation. Quenching of two soliton states or stripe solitons in the attractive case, discussed in section \ref{sec:5:1}, produces qualitatively similar results as that of repulsive SO coupled BEC, as illustrated in figures \ref{fig10}, \ref{fig13}, and \ref{fig15}.

\section{Summary}
\label{sec:7}

In this paper, we have analyzed the dynamics of spin-orbit coupled Bose-Einstein condensates with Rabi mixing in a quasi one-dimensional setting by numerically solving coupled Gross-Pitaevskii equations.

First, we reported the results on the quench induced dynamics over stationary configurations due to sudden changes in the spin-orbit and Rabi coupling parameters for various intra- and inter-species interaction strengths. We studied the dynamics of Bose-Bose bright solitons in SO coupled BECs with attractive intra- and inter-species interactions. By preparing a stationary profile of the wavefunctions from the numerical solutions of the coupled Gross-Pitaevskii equations, we examined the stability and dynamics of the solitons through time evolution. The stationary profile is maintained during the time evolution provided the initial wavefunctions, $\psi_\uparrow(x) $ and $ \psi_\downarrow(x)$, obey certain anti-symmetry property, that is, $\psi_\uparrow(x) = - \psi_\downarrow(-x)$. On the other hand, the solitons propagate when the initial wavefunctions possess a cross-symmetry of the form $\psi_\uparrow(x) = \psi_\downarrow(-x)$. We noticed that the symmetry properties of the stationary wavefunctions influence the soliton dynamics significantly.

Next, we studied the dynamics of spin-orbit coupled BECs by suddenly modifying the SO and Rabi coupling parameters during time evolution. For attractive interactions, in a binary BEC, quenching of spin-orbit coupling shows decay and revival phenomenon of bright soliton. Whereas the simultaneous introduction of $k_{L}$ and $\Omega$ also leads to a similar dynamics with oscillations. However, in spin-orbit coupled BECs, sudden change in $\Omega$ during time evolution results in the segregation of the initially spin-mixed state into two parts along with breather like patterns, and these parts move away from each other as time progresses. The inherent lack of Galilean invariance in SO coupled BECs due to spin-orbit interaction gives rise to the shape-changing phenomenon of solitons for the attractive interactions. Further, we observed that for relatively higher values of $\Omega$, the shape-changing effect gets suppressed. For repulsive interactions, quenching of the coupling parameters exhibit a large variety of phenomena such as breather-like oscillations, spin mixing-demixing dynamics, miscible-immiscible transition, dark-bright solitons, multiple dark-soliton dynamics, and nonlinear wave patterns. We also noticed that, the time at which the quench is applied has an effect on the number of dark-bright solitons.

In the present study, we demonstrate that spin transport dynamics can be controlled by suitably modifying the coupling strengths. This may be of importance, for instance, in the manipulation of spin-orbit interactions in materials which furnish tunable spin qubit~\cite{Awschalom2012,Petersson2012}. Spin-orbit coupled BEC is a natural candidate for spin-qubits due to the double degeneracy linked with the pseudo-spin degree of freedom~\cite{Stanescu2008}. By sudden switching of coupling strengths on the condensate, we show a variety of spin-dynamics. The phase factor added to the spin components due to the presence of SO coupling drives mechanical motion and the spin-flip transition. Also, the coherent singlet state can be split (demixed) and then recombined (mixed) as time progresses due to the spin-rotation accomplished by the spin-orbit coupling. Such a coherent superposition of spin-qubits will be useful in quantum computation and information processing~\cite{Mardonov2015}.

The quenching of spin-orbit coupled BEC that leads to demixing and mixing dynamics of the spin states could be of use to dynamically revive the spin-states or spin qubits. In this paper, the condensate parameters are chosen so as to fall within the experimentally feasible range of $^{39}$K BEC~\cite{Jin2014,Roati2007}. The present study would be appropriate for understanding the quench dynamics of spin-orbit coupled Bose-Einstein condensates of $^{39}$K atoms. 

\appendix

\section{Quasi-BEC regime: One-dimensional model}
\label{app:a}
According to Refs.~\cite{Luca2004,Luca2005}, at zero temperature a system of dilute bosons of mass $m$ and under transverse harmonic confinement of frequency $\omega_{\bot}$, lives in the one-dimensional (1D) quasi-BEC regime under the conditions 
\begin{equation}
{a_s\over a_{\bot}^2} \ll \rho_{1d}(x,t) \ll {1\over a_s} \; , 
\label{crucial}
\end{equation}
where $a_s$ is the 3D s-wave scattering length of the interaction, $a_{\bot}=\sqrt{\hbar/(m\omega_{\bot})}$ is the characteristic length of the transverse harmonic confinement of frequency $\omega_{\bot}$, and $\rho_{1d}(x,t)$ is the local 1D axial density. 

Under the conditions (\ref{crucial}) the bosonic system is well described by the 1D GP equation 
\begin{equation}
\mathrm{i} \hbar \partial_t \psi_{1d}(x,t) = \left[ -{\hbar^2\over 2m}\partial_x^2 + g_{1d} |\psi_{1d}(x,t)|^2 \right] \psi_{1d}(x,t) 
\end{equation}
where $\psi_{1d}(x,t)$ is the axial wavefunction, such that 
\begin{equation}
\rho_{1d}(x,t) = |\psi_{1d}(x,t)|^2 \; , 
\end{equation}
and 
\begin{equation}
g_{1d} = {g_{3d}\over 2\pi a_{\bot}^2} = {2\hbar^2 a_s\over m a_{\bot}^2}
\end{equation}
is the 1D interaction strength with $g_{3d}$ the 3D interaction strength. 
Notice that 
\begin{equation}
\int_{-\infty}^{+\infty} \rho_{1d}(x,t) dx = N 
\end{equation}
with $N$ being the total number of bosons in the 1D quasi-BEC. 

It is convenient to use the scaled variables, ${\tilde x}= {x/ a_{\bot}}$, ${\tilde t} = \omega_{\bot} t $, ${\tilde \psi}_{1d} = \sqrt{a_{\bot}} \psi_{1d}$, and ${\tilde \rho}_{1d} = a_{\bot} \rho_{1d}$.
In this way the 1D GP equation becomes 
\begin{equation}
\mathrm{i} \partial_{\tilde t} {\tilde \psi}_{1d}({\tilde x},{\tilde t}) = \left[ -{1\over 2}\partial_{\tilde x}^2 + {\tilde g}_{1d} |{\tilde \psi}_{1d}({\tilde x},{\tilde t})|^2 \right] {\tilde \psi}_{1d}({\tilde x},{\tilde t}) \; , 
\end{equation}
where 
\begin{equation}
{\tilde g}_{1d} = 2 {a_s\over a_{\bot}} \; . 
\label{u2}
\end{equation}
The conditions (\ref{crucial}) can be obviously re-written as 
\begin{equation}
{a_s\over a_{\bot}} \ll a_{\bot }\rho_{1d}(x,t) \ll {a_{\bot}\over a_s} \; . 
\end{equation}
Taking into account the scaling, we then obtain 
\begin{equation}
{1\over 2} {\tilde g}_{1d} \ll {\tilde \rho}_{1d}({\tilde x},{\tilde t}) \ll {2\over {\tilde g}_{1d}} \; 
\end{equation}
or, equivalently 
\begin{equation}
{1\over 4} {\tilde g}_{1d}^2 \ll {{\tilde g}_{1d} \over 2} \ {\tilde \rho}_{1d}({\tilde x},{\tilde t}) \ll 1 \; . 
\label{q1dregion}
\end{equation}
Based on the above condition (\ref{q1dregion}), we examine the validity of the quasi-condensate limit for the interaction parameters considered in the present work. We used the quasi-1D condensate model with the repulsive scattering lengths ranging from $3.785 a_0$ to $118.25 a_0$. Taking into account the maximum of the computed 1D density, $\tilde\rho_{1d}({\tilde x},{\tilde t})$ and $a_\perp = 1 \,\mu\mbox{m}$, in the lower limit of the scattering length ($a_s = 3.785 a_0$), the above condition turns out to be $4 \times 10^{-8} \ll 9 \times 10^{-6} \ll 1$. On the other hand, it becomes $3.9 \times 10^{-5} \ll 1.15 \times 10^{-4} \ll 1$ when $a_s = 118.25 a_0$ (upper limit). Also, it is easy to argue that the BEC lies in the quasi-condensate regime as long as ${{\tilde g}_{1d}} \ {\tilde \rho}_{1d}({\tilde x},{\tilde t}) /2 \ll 1$ is valid~\cite{Luca2004,Luca2005}.

\section{Symbiotic stripe soliton in SO coupled BECs with uneven intra-species interactions}
\label{app:b}
It is worth investigating the dynamics of stripe soliton under uneven intra-species interactions of the spin components. As pointed out in section \ref{sec:4}, we identify a symbiotic stripe soliton by fixing the SO coupling strength as $k_L = 2$ and Rabi coupling strength as $\Omega =1$, attractive inter-species interaction of $\beta = -0.8$, and repulsive intra-species interactions of $\alpha_{\uparrow \uparrow} = 0.8$, and $\alpha_{\downarrow \downarrow} = 0.2$. %
\begin{figure}[!ht]
\begin{center}
\includegraphics[width=0.99\linewidth]{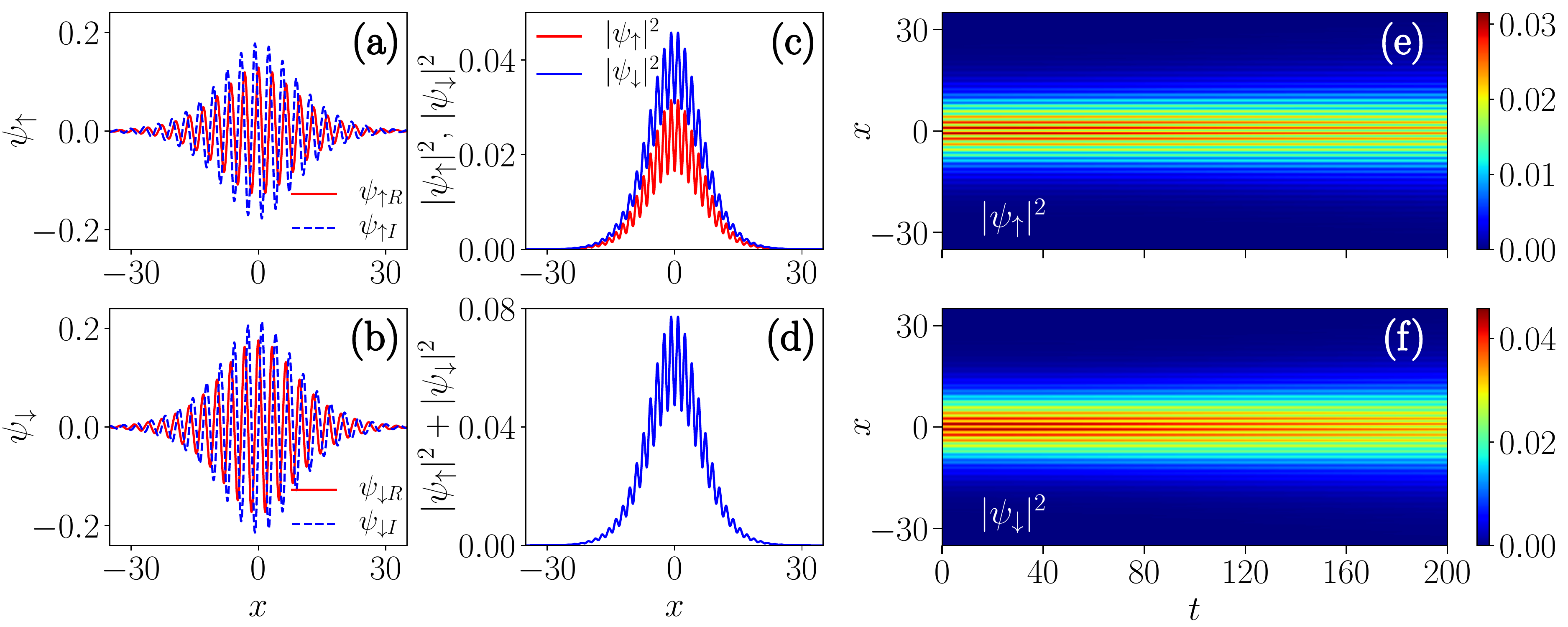}
\end{center}
\caption{Plots of real (solid red line) and the imaginary (dashed blue line) parts of the stationary profiles (a) $\psi_{\uparrow}(x)$ and (b) $\psi_{\downarrow}(x)$ of the symbiotic stripe soliton for $\lambda = 0$, $\alpha_{11} = 0.8, \alpha_{22}= 0.2, \beta = -0.8$, $k_L = 2.0$, and $\Omega = 1.0$ in equation (4) with $\mu_{\uparrow} = \mu_{\downarrow} \approx -0.886 $. (c) The spin densities $\vert \psi_{\uparrow} \vert ^2$ (red solid line) and $\vert \psi_{\downarrow} \vert ^2$ (solid blue line), and (d) the total density $\vert \psi_{\uparrow} \vert ^2 + \vert \psi_{\downarrow} \vert ^2$. The time evolution of the spin densities are shown in (e) and (f).}
\label{fig-Ab}
\end{figure}%
Figures~\ref{fig-Ab}(a) and \ref{fig-Ab}(b) show the stationary wavefunctions of the spin components obtained using imaginary time propagation. In figure \ref{fig-Ab}(c), we plot the densities of the spin components, which reveal the symbiotic nature of the stripe solitons. One may notice that the spin-up component denoted by the red line in figure \ref{fig-Ab}(c) has a lower number of atoms while the spin-down component (blue line) is populated with a larger fraction. Figure \ref{fig-Ab}(d) depicts the total density. The time evolution of these symbiotic stripe solitons are shown in  figures~\ref{fig-Ab}(e) and \ref{fig-Ab}(f). We observed that the symbiotic nature is preserved during the time evolution. Further, we found that the symbiotic nature is also present in stripe solitons exhibited by a fully attractive SO coupled BEC. However, attractive intra- and repulsive inter-species interactions do not hold the symbiotic stripe solitons while it evinces symbiotic bright-bright solitons.

\ack
RR acknowledges the University Grants Commission (UGC), India for financial support in the form of UGC-BSR-RFSMS Research Fellowship. TS acknowledges financial support by Council of Scientific and Industrial Research (CSIR), India under Grant No. 03(1422)/18/EMR-II. The work of PM forms parts of sponsored research projects by Council of Scientific and Industrial Research (CSIR), India under Grant No. 03(1422)/18/EMR-II, and Science and Engineering Research Board (SERB), India under Grant No. CRG/2019/004059. LS acknowledges partial support from the DOR-BIRD 2019 Project ``Struttura della Materia'' of the University of Padova.

\end{document}